\documentclass[prb,twocolumn,superscriptaddress,showpacs,amsmath,amssymb,longbibliography,floatfix]{revtex4-2}
\usepackage{amsmath, amsfonts, amssymb, bm}
\usepackage{float}
\usepackage{mathtools}
\usepackage[inline]{enumitem}
\usepackage{graphicx}
\usepackage{xcolor}
\usepackage[breaklinks,hypertexnames=false]{hyperref}
\hypersetup{colorlinks,linkcolor={magenta},citecolor={blue},urlcolor={blue}}
\usepackage[capitalize]{cleveref}

\begin{document}
	\title{Engineering subgap states in superconductors by the symmetry of altermagnetism} 
	
	\newcommand{\tianjin}{Department of Physics, Tianjin University, Tianjin 300354, China}
	
	\newcommand{\nagoya}{Department of Applied Physics, Nagoya University, Nagoya 464-8603, Japan}
	
	\newcommand{\Okayama}{Faculty of Environmental Life, Natural Science and Technology, Okayama University, 700-8530 Okayama, Japan}
	
	\newcommand{\Uppsala}{Department of Physics and Astronomy, Uppsala University, Box 516, S-751 20 Uppsala, Sweden}
	
	\newcommand{\uam}{Department of Theoretical Condensed Matter Physics, Universidad Aut\'onoma de Madrid, 28049 Madrid, Spain}
	\newcommand{\ifimac}{Condensed Matter Physics Center (IFIMAC), Universidad Aut\'onoma de Madrid, 28049 Madrid, Spain}
	\newcommand{\inc}{Instituto Nicol\'as Cabrera, Universidad Aut\'onoma de Madrid, 28049 Madrid, Spain}

    \author{Bo Lu}
    \affiliation{\tianjin}

    \author{Phillip Mercebach}
	\affiliation{\uam}
	\affiliation{\ifimac}
	
	\author{Pablo Burset}
	\affiliation{\uam}
	\affiliation{\ifimac}
	\affiliation{\inc}

        \author{Keiji Yada}
	\affiliation{\nagoya}
    
	\author{Jorge Cayao}
	\affiliation{\Uppsala}
	
	\author{Yukio Tanaka}
	\affiliation{\nagoya}

       \author{Yuri Fukaya}
	\affiliation{\Okayama}

	\date{\today}
	%\linenumbers

\begin{abstract}
Combining superconducting and magnetic materials is a promising path to generate exotic interface subgap states.  
In this regard, altermagnetism is particularly interesting because it lifts spin degeneracy while providing tailored anisotropy of spin splittings. 
Here, we investigate the realization and control of subgap states by using the symmetry contrast between altermagnetic fields and unconventional pairings.
When the symmetries of altermagnetism and unconventional superconductivity align, we demonstrate the emergence of bulk zero-energy flat bands as the Bogoliubov Fermi surface, giving rise to a zero-bias conductance peak. 
The symmetry and strength of $d$-wave altermagnets strongly affect the surface Andreev states from $d$-wave and chiral $d$- and $p$-wave superconductors. As a result, distinct types of subgap states are realized, including curved and flat bands, that can be detected by tunneling spectroscopy. 
% Furthermore, we find that the altermagnetism-induced subgap states give rise to a large spin conductance at zero net magnetization which helps identify the strength of the underlying altermagnetism and superconductivity. 
Our results offer a solid route for designing and manipulating subgap states in superconducting systems, which can be useful for functionalizing superconducting devices. 
\end{abstract}

\maketitle

\emph{Introduction.}---
The ability to induce and control subgap states in superconducting systems is a central topic in condensed matter physics, with a profound fundamental relevance and high potential for quantum applications. One example of such subgap states is the surface Andreev bound states (SABSs)~\cite{Hu94,TK95,Covington97,Alff97,Strasik98,kashiwaya2000,Lofwander2001,Ryu002,Msato11,tanakaJPSJ12,Mizushima18,tanaka2024review,asano2021andreev}, which emerge due to the anisotropic nature of the superconducting pair potentials and can be detected by tunneling spectroscopy~\cite{kashiwaya2000}. Interestingly, SABSs in unconventional superconductors (SCs) can lead to Majorana states~\cite{SatoAndo2017,cayao2019odd,doi:10.7566/JPSJ.85.072001,tanaka2024review}, which hold promise for designing topological quantum bits~\cite{Nayakrev08,Sarmarev2015,beenakker2020search,Aguado2020,Oreg2020,Aguadoapl2020}. 
Moreover, certain superconducting setups can also host bulk subgap states, such as Bogoliubov-Fermi surfaces (BFSs)~\cite{Brydonreview,Pablo15,PhysRevB.96.094526,PhysRevLett.118.127001,Zhu_2021,PhysRevB.107.104515,PhysRevB.97.115139,PhysRevB.102.064504,PhysRevB.110.245417,pal2024transport}, which signal the nontrivial topological nature of superconductivity~\cite{PhysRevLett.128.107701,Zhu_2021}. 
For practical purposes, subgap states need to be manipulated, often carried out by coupling external fields to their intrinsic quantum numbers~\cite{de2010hybrid,Prada2020,flensberg2021engineered}, such as orbital or spin. Although this can indeed be achieved by ferromagnets~\cite{Ranran2023,MESERVEY1994173,Yokoyama07}, their unavoidable finite magnetization and stray fields make them detrimental for superconductivity \cite{Paschoa2020}, and therefore, other materials are required. 

\begin{figure}[!t]
\begin{center}
\includegraphics[width=86mm]{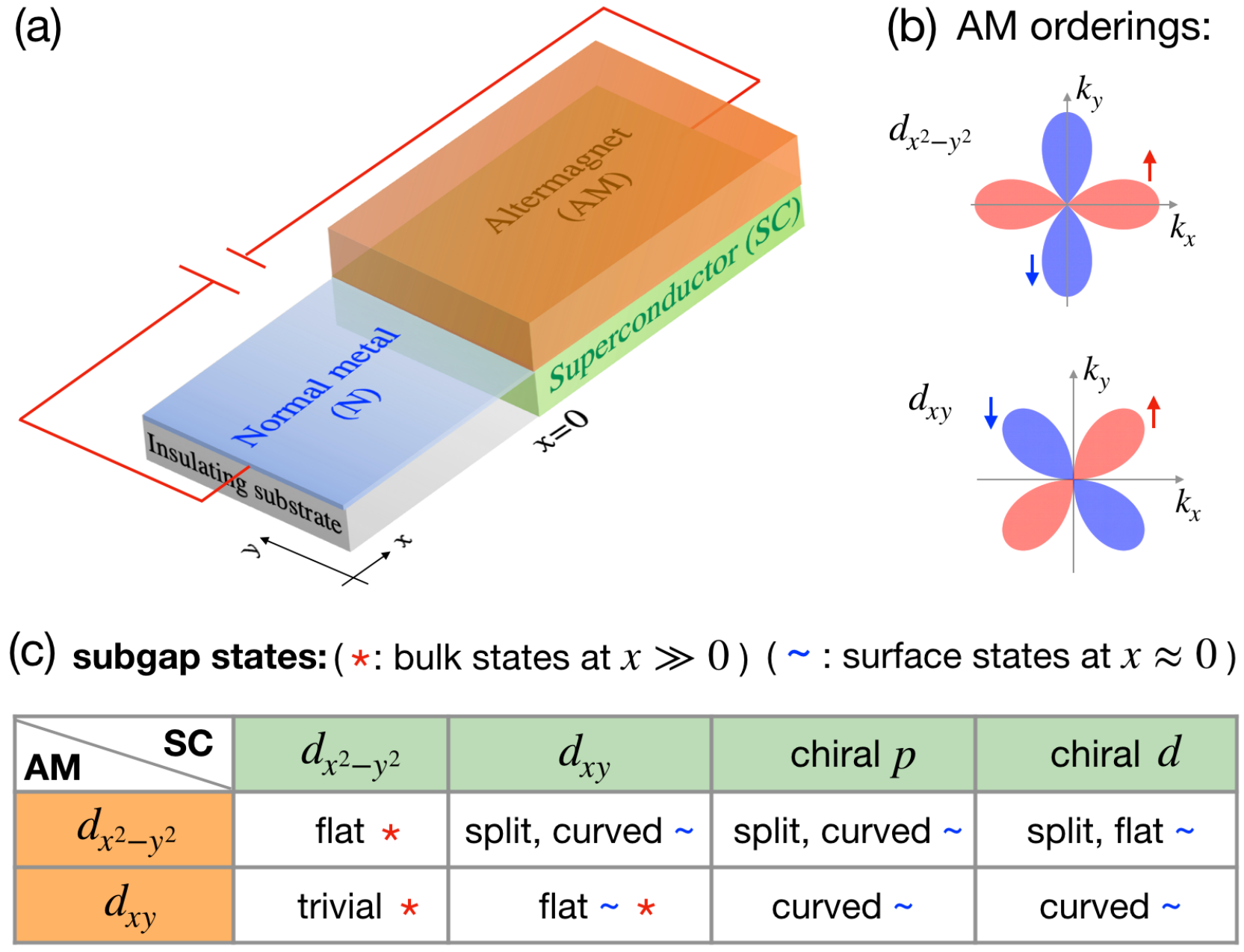}
\end{center}
\caption{(a) An altermagnet (AM, orange) is placed on top of a superconductor (SC, green) to engineer subgap states. A normal metal probe (N, blue) is coupled to SC at $x=0$ near the AM-SC interface, achieved by adjusting the thickness of the insulating substrate (grey) beneath N. 
(b) Illustrations of the selected AM orderings in momentum space where the red and blue colors indicate up and down spins. (c) Table listing the type of emergent subgap states for several AM-SC combinations. Labels ``flat'', ``split'', and ``curved'' indicate when the dispersion of subgap states is respectively zero, spin-split, or drastically varying.}
\label{Fig1}
\end{figure}

With the recent advent of altermagnetism \cite{Libor22,landscape22}, a powerful avenue has emerged for coupling spins in superconductors thanks to the altermagnetic spin-orbit properties~\cite{Hayami19,NakaNatCommun2019,Hayami20,NakaPRB2020,Ma2021,Qihang22,Mazin22,Bailing,krempasky2024,Song2025,fukayarev}. Altermagnets (AMs) are spin-compensated magnetic materials with a nonrelativistic momentum-dependent spin splitting and anisotropic spin-polarized Fermi surfaces~\cite{Wu2004}, which result in zero net magnetization and hence vanishing stray fields unlike ferromagnets~\cite{Giilmemory,Giil2024}. In combination with SCs, AMs have already been shown to host intriguing properties~\cite{fukayarev}, such as orientation-dependent Andreev reflections~\cite{Sun23,Papaj23,maeda24,Niu_2024,nagae2024,Wenjun25}, anomalous Josephson currents~\cite{zhang2024,Ouassou23,Beenakker23,Bo2024,fukaya2024,sun2024,Cheng24}, nontrivial light-matter coupling~\cite{fu2025floquet,fu2025light}, superconducting diodes~\cite{Banerjeediode04,qingfengdiode24,perfectdiode25}, 
spin-triplet Cooper pairs~\cite{Maeda2025, chakraborty2024,khodas2025strain,PhysRevB.111.054520,parshukov2025}, Majorana states~\cite{cano23,Zhongbo23,CCLiu1,CCLiu2,mondal2024,hadjipaschalis2025Mzm}, among other phenomena~\cite{Carvalhoprb2024,Sukhachov24,SeungBeom25,Hujinxin25,sumita2025phase,Maiani25,Belzig25,chatterjee2025interplay,Yokoyama25floquet,hodge2025platform,KokkelerSciPost25}. 
Despite the impressive advances, the use of altermagnetism for realizing subgap states is still an open problem. 

In this work, we demonstrate the emergence and control of highly tunable subgap states by using altermagnetism in unconventional SCs. To show this, we focus on a hybrid system composed of a two-dimensional SC with altermagnetism due to its proximity to an AM [Fig.\,\ref{Fig1}(a)], where a normal probe enables the detection of bulk and surface subgap states. We consider various types of unconventional SCs combined with $d$-wave AMs [Fig.\,\ref{Fig1}(b)], which allows us to reveal different types of subgap states, including bulk zero energy flat bands and SABSs, as listed in Fig.\,\ref{Fig1}(c). Such subgap states are driven by the symmetry contrast between competing superconducting and magnetism orders and give rise to strong charge conductance signatures, enabling their detection and also the identification of the type of altermagnetism supporting them. 

\emph{Model.}---
We aim to harness altermagnetism to realize and detect subgap states in unconventional SCs. 
We thus consider a bilayer AM-SC device where the coupling between AM and SC accounts for a proximity-induced AM field in the SC. Then, a two-dimensional normal metal lead N is connected to the SC at $x=0$ to probe the  properties of altermagnetic superconductors, see Fig.\,\ref{Fig1}(a). 
The N-SC planar junction
is modeled by $\mathcal{H}=(1/2)\sum\nolimits_{\bm{k}}\hat{c}_{%
\bm{k}}^{\dag }\mathcal{H}_{\bm{k}}\hat{c}_{\bm{k}}$, where the Nambu spinor is given by $\hat{c}_{\bm{k}}=(c_{\bm{k}\uparrow },c_{%
\bm{k}\downarrow },c_{-\bm{k}\uparrow }^{\dag },c_{-\bm{k}\downarrow }^{\dag
})^{T}$, $\bm{k}=(k_{x},k_{y})$, and 
\begin{equation}
\label{Eq1}
    \mathcal{H}_{\bm{k}}=\left[ \xi_{\bm{k}} +U(x) \right]
    \hat{\tau}_{z}+\mathcal{M}_{\bm{k}}(x)\hat{s}_{z}\hat{\tau}_{z}+\Delta_{\bm{k}}(x).
\end{equation}
Here, $\xi_{\bm{k}}=\hbar^{2}|\bm{k}|^{2}/(2m)-\mu$ is the kinetic energy, with $\mu$ being the chemical potential, $U(x)=U_{0}\delta_{x}$ is the interface barrier potential at $x=0$, with $U_{0}$ the barrier strength defining the   parameter $Z=mU_{0}/(\hbar^{2}k_{F})$ that characterizes the tunneling ($Z\gg1$) and full transparent ($Z=0$) regimes, $k_{F}=\sqrt{2m\mu/\hbar^{2}}$, and $\hat{s}_{i}( \hat{\tau}_{i}) $ is the $i$-th Pauli matrix in spin (Nambu) space. 
Moreover, $\mathcal{M}_{\bm{k}}(x)=\Theta(x)h_{0}m_{\bm{k}}$ is the altermagnetic field, with $h_{0}$ denoting the altermagnetic strength and $m_{\bm{k}}$ the anisotropic factor, and, without loss of generality, the Neel vector of the AM is chosen along the $z$-axis.  The unconventional superconducting pair potential $\Delta_{\bm{k}}(x)=\Theta(x)\hat{\Delta}_{\bm{k}}$ is only finite in the SC. We assumed $h_{0}\ll \mu $ so that the wavevectors at low energies can be approximated by $\bm{k}=k_{F}\bm{\hat{k}}$, with the unit vector $\bm{\hat{k}}=(\hat{k}_{x},\hat{k}_{y})$. 

We consider highly relevant AMs exhibiting $d$-wave
parity [Fig.\,\ref{Fig1}(b)] 
expected in, e.g., RuO$_{2}$ and KV$_{2}$Se$_{2}$~\cite{landscape22,Jiang2025}, 
and modeled by specific anisotropic factors $m_{\bm{k}}$. We have 
$m_{%
\bm{k}}=(\hat{k}_{x}^{2}-\hat{k}_{y}^{2})\cos 2\alpha +2\hat{k}_{x}\hat{k}%
_{y}\sin 2\alpha $,
with $\alpha=0$ for $d_{x^{2}-y^{2}}$-wave AMs and $\alpha =\pi /4$ for $d_{xy}$-wave AMs. 
For the unconventional SCs, we consider four types of experimentally relevant pairings~\cite{RevModPhys.63.239,tanaka2024review}. For $d$-wave SCs, we have
$\hat{\Delta}_{\bm{k}}=-\Delta _{0}s_{\bm{k}}\hat{s}_{y}\hat{\tau}_{y}$,
with $s_{\bm{k}}=(\hat{k}_{x}^{2}-\hat{k}_{y}^{2})\cos 2\chi +2\hat{k}_{x}%
\hat{k}_{y}\sin 2\chi $. Here, i) $\chi =0$ represents $d_{x^{2}-y^{2}}$-wave pairing and ii) $\chi =\pi /4$ $d_{xy}$-wave. These two pairing states are related by a rotation with respect to the orientation of the interface with the metallic probe. 
Chiral SCs display iii) chiral
$p$-wave pairing $\hat{\Delta}_{\bm{k}}=\Delta _{0}\hat{s}%
_{x} ( \hat{k}_{x}\hat{\tau}_{x}-\hat{k}_{y}\hat{\tau}_{y} ) s_{%
\bm{k}}$ or iv) chiral $d$-wave pairing $\hat{\Delta}_{\bm{k}%
}=-\Delta _{0} ( \hat{k}_{x}^{2}-\hat{k}_{y}^{2} ) \hat{s}_{y}\hat{%
\tau}_{y} s_{\bm{k}}-2\Delta _{0}\hat{k}_{x}\hat{k}_{y}\hat{s}_{y}\hat{\tau}%
_{x}s_{\bm{k}}$. In both cases $s_{\bm{k}}=1$. Here, $\Delta_0$ is the gap amplitude in the presence of the AM field and can be suppressed~\cite{Belzig25,SM}.

The quasiparticle energy dispersion in the SC with altermagnetism is given by 
\begin{equation}
\label{dispersion}
E_{\bm{k}\nu}^{\pm}=\nu h_{0}m_{\bm{k}}\pm\sqrt{\xi_{\bm{k}}^{2}+|\Delta_0 s_{\bm{k}
}|^{2}}\,,
\end{equation}
where $\nu=\pm$ indicating two possible anisotropic energy shifts ($\pm h_{0}m_{\bm{k}}$) of subbands. From Eq.\,(\ref{dispersion}), it is easy to see that, for $\nu h_{0}m_{\bm{k}}=|\Delta_0 s_{\bm{k}}|$, we get $E^{+}_{\bm{k}\nu}\neq0$ and $E^{-}_{\bm{k}\nu}=0$, the latter signaling a zero energy bulk flat band akin to BFSs~\cite{Brydonreview}. 
This is an example of a \emph{bulk} subgap state induced by altermagnetism, and its detection can be carried out via conductance in the setup of Fig.\,\ref{Fig1}(a). Notably, this setup also allows us to search for altermagnetism-induced  subgap states at the \emph{surface}, which can be directly accessed via conductance as we explain below.

To obtain the conductance, we employ the scattering approach, which is based on the wavefunctions of injected electrons with spin $\sigma=\uparrow,\downarrow$. Due to translational invariance along  $y$, the wavefunctions are $\Psi
_{\sigma}(x,k_{y})=\psi _{\sigma}(x)e^{ik_{y} y}$. In the normal side, $\psi _{\sigma}(x)$ is given by%
\begin{equation}
\psi _{\sigma}(x<0)=[\hat{A}_{\sigma}+a_{\sigma}\hat{B}_{\sigma}]
e^{ik_{x}x}+b_{\sigma}\hat{A}_{\sigma}e^{-ik_{x}x}\,,
\end{equation}%
where $\hat{A}_{\uparrow }=\left( 1,0,0,0\right) ^{T}$, $\hat{A}_{\downarrow
}=\left( 0,1,0,0\right) ^{T}$, $\hat{B}_{\downarrow }=\left( 0,0,1,0\right)
^{T}$ and $\hat{B}_{\uparrow }=\left( 0,0,0,1\right) ^{T}$, $k_{x}=k_{F}\cos
\theta $, and $k_{y}=k_{F}\sin \theta $. The wave functions in the SC can be
written in a similar way. Then, the coefficients $a_{\sigma }$ and $%
b_{\sigma }$ are obtained from the boundary conditions $\psi (0^{+})=\psi
(0^{-})$, $\partial _{x}\psi (0^{+})-\partial _{x}\psi
(0^{-})=(2mU_{0}/\hbar ^{2})\psi (0)$. The angle-resolved charge conductance
is found as \cite{BTK82} $\bar{\sigma}_{\sigma }^{e}=1+|a_{\sigma
}|^{2}-|b_{\sigma }|^{2}$, which gives
\begin{equation}
\bar{\sigma}_{\sigma }^{e}=\bar{\sigma}_{n}\frac{1+\bar{\sigma}_{n}|\Gamma
_{\sigma }^{+}|^{2}+(\bar{\sigma}_{n}-1)|\Gamma _{\sigma }^{+}\Gamma
_{\sigma }^{-}|^{2}}{|\Lambda -\eta (1-\bar{\sigma}_{n})\Gamma _{\sigma
}^{+}\Gamma _{\sigma }^{-}|^{2}}\,,  \label{KT}
\end{equation}%
with $\bar{\sigma}_{n}=1/[1+Z^{2}/(\cos ^{2}\theta )]$. Here, $\Gamma
_{\sigma }^{\pm }$ are given by
\begin{equation}
\Gamma _{\sigma }^{\pm }=\frac{\Delta _{0}s_{\bm{k}_{\pm }}}{E_{\sigma ,%
\bm{k}_{\pm }}+\sqrt{E_{\sigma ,\bm{k}_{\pm }}^{2}-\Delta _{0}^{2}s_{\bm{k}%
_{\pm }}^{2}}}.  \label{Gamma}
\end{equation}%
where $E_{\uparrow ,\bm{k}_{\pm }}=E-\mathcal{M}_{\bm{k}_{\pm }}$ and $%
E_{\downarrow ,\bm{k}_{\pm }}=E+\mathcal{M}_{\bm{k}_{\pm }}$, with $\bm{k}%
_{+}$, $\bm{k}_{-}$ are given by $\bm{k}_{\pm }=k_{F}\left( \pm \cos \theta
,\sin \theta \right) $. Moreover, $\Lambda =1$ for $d_{x^{2}-y^{2}}$- and $%
d_{xy}$-wave SCs; $\Lambda =e^{2i\theta }$ for chiral $p$-wave SCs; $\Lambda
=e^{4i\theta }$ for chiral $d$-wave SCs. Also, $\eta =1$ for spin-singlet
SCs (cases i,ii,iv), while $\eta =-1$ for spin-triplet SCs (case iii).
Hence, the charge conductance in the superconducting state is calculated as
\begin{equation}
\sigma _{\mathrm{SC}}^{e}=\frac{e^{2}}{h}\int_{-\frac{\pi }{2}}^{\frac{\pi }{%
2}}\bar{\sigma}^{e}\cos \theta d\theta ,  \label{esconductance}
\end{equation}%
where $\bar{\sigma}^{e}=\bar{\sigma}_{\uparrow }^{e}+\bar{\sigma}%
_{\downarrow }^{e}$ is the angle-resolved charge conductance. Later we
normalize the conductance by $\sigma _{N}=(2e^{2}/h)\int \bar{\sigma}%
_{n}\cos \theta d\theta $.

At this point, it is important to highlight that, to access the surface ABSs states via conductance, the necessary condition is to set $\bar{\sigma} _{n}\rightarrow 0$ and find the zeroes of the denominator in Eq.\,(\ref{KT}), which leads to
\begin{equation}
\label{condition}
\Gamma _{\sigma}^{+}\Gamma_{\sigma}^{-}=\Lambda /\eta\,.
\end{equation}
This condition corresponds to the poles of the associated scattering matrix. Thus, the surface ABSs are obtained from Eq.\,(\ref{condition}). In what follows, we explore how the bulk subgap state discussed below Eq.\,(\ref{dispersion}) as well as surface ABSs can be induced by altermagnetism and detected by conductance in unconventional SCs. Our results in the continuum model are the same with the lattice model shown in the Supplementary Material \cite{SM}. 

\begin{figure}[!t]
\begin{center}
\includegraphics[width=88mm]{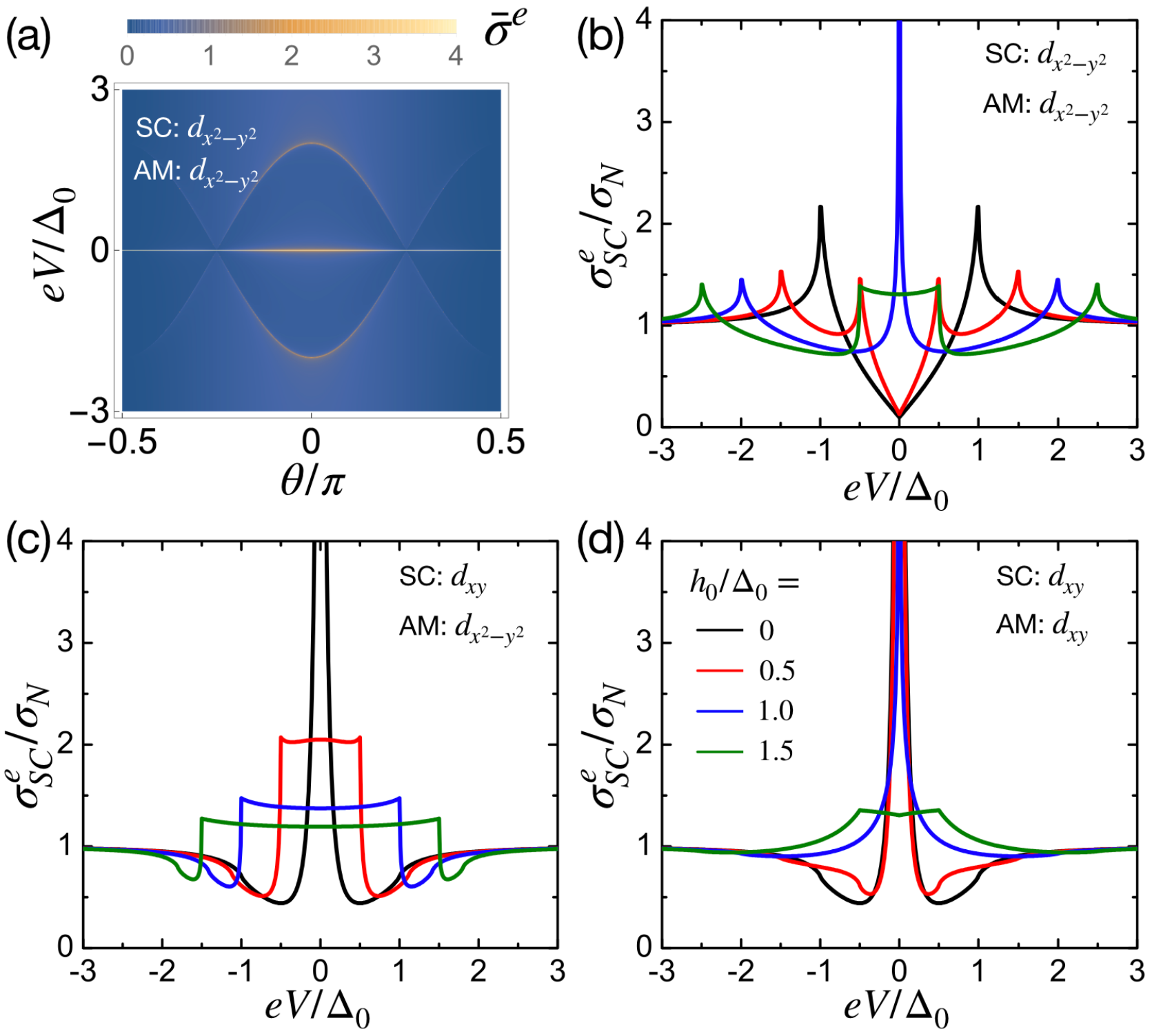}
\end{center}
\caption{Conductance for a junction with various combinations of $d$-wave SCs and $d$-wave AMs. 
%Panels (a),(b) for  $d_{x^{2}-y^{2}}$-wave SC,  (c),(d) for $d_{xy}$-wave SC, (a), (c) for $d_{x^{2}-y^{2}}$-wave AM and (b), (d) for $d_{xy}$-wave AM. 
The AM is $d_{x^{2}-y^{2}}$-wave in panels (a-c) and $d_{xy}$-wave in (d). SC pairing is $d_{x^{2}-y^{2}}$-wave in (a), (b) and $d_{xy}$-wave in (c), (d). 
Panel (a) show angle-resolved conductance at $h_{0}=\Delta_{0}$. For all panels $Z=2$.} 
\label{Fig2}
\end{figure}

\emph{Bulk and surface flat bands by $d$-wave AMs in $d$-wave SCs.}---
We begin by analyzing the junction formed by $d_{x^{2}-y^{2}}$-wave SC
with $d$-wave altermagnetism. Before delving into the conductance, we remind
our discussion below Eq.\thinspace (\ref{dispersion}), where we show that at
$\nu h_{0}m_{\bm{k}}=|\Delta _{0}s_{\bm{k}}|$, one of the bands reaches zero
energy, namely, $E_{\bm{k}\nu }^{-}=0$, while $E_{\bm{k}\nu }^{+}\neq 0$.
Thus, when the AM and SC have the same symmetry ($m_{\bm{k}}=s_{\bm{k}}$) a
zero-energy flat band emerges as a \emph{bulk} subgap state entirely induced
by altermagnetism and akin to BFSs. 
Indeed, this flat band is only possible due to the accidental disappearance of a gaped spectrum at $E=0$ while the superconducting gap still exists at $|E|<2\Delta_0$, maintaining superconducting coherence.
As a result, the bulk density of states
at zero energy becomes singular, similar to the divergent density of states
at $E=\Delta _{0}$ for a conventional $s$-wave SC. This effect, and hence
the flat band, can be detected via conductance, as we demonstrate in
Fig.\thinspace \ref{Fig2}, where we plot the charge conductance showing nontrivial subgap states with $d$-wave AMs and $d$-wave SCs. The first feature to notice is that, for a $d_{x^2-y^2}$-wave SC with $d_{x^2-y^2}$-wave AM, the angle-resolved charge conductance $\bar{\sigma}^{e}(eV)$ at $h_{0}=\Delta _{0}$ is greatly
enhanced at $eV=0$ as a result of the BFS ($E_{\bm{k}\nu }^{-}=0$) discussed
above [Fig.\thinspace \ref{Fig2}(a)]. This enhancement gives rise to a zero-bias conductance peak (ZBCP) in the total charge conductance $\sigma _{\mathrm{SC}%
}^{e}(eV)$ at $h_{0}=\Delta _{0}$ [Fig.\thinspace \ref{Fig2}(b)]. The ZBCP stems from bulk BFSs and thus its width is very narrow~\cite{fukaya2025crossed}. By moving away from $eV=0$ and for $h_{0}\neq \Delta _{0}$, $\sigma _{\mathrm{SC}}^{e}(eV)$ develops a
V-shape profile with spikes outside the gap, an effect that is similar to ferromagnetic SCs \cite{Yokoyama07}. 
It is worth mentioning that when $d_{xy}$-wave
AM and $d_{x^{2}-y^{2}}$-wave SC coexist, no zero energy bulk flat band can be found according to Eq.\thinspace (\ref{dispersion}).

Now we turn to $d_{xy}$-wave SCs. It is well-known that $d_{xy}$-wave pair
potentials lead to zero-energy surface flat states without AMs \cite%
{tanaka2024review}. Here we are interested in analyzing how AM can control
the surface states and how to induce them in regimes not previously present.
This is demonstrated in Figs.\thinspace \ref{Fig2}(c,d), where we plot the total charge conductance. For a $d_{x^{2}-y^{2}}$-wave AM and
$d_{xy}$-wave SC, 
% the angle-resolved charge conductance $\bar{\sigma}^{e}$
% as a function of $\theta $ shows that the surface states split in a curved
% shape, with the size of the splitting given by the anisotropic AM field, but
% remain degenerate at $\theta =\pi /4$ [the inset in Fig.\thinspace \ref{Fig2}%
% (c)]. Due to this reason,
the total charge conductance $\sigma _{\mathrm{SC}%
}^{e}$ gets suppressed as the AM field increases, going from a sharp ZBCP at
$h_{0}=0$ into an almost constant subgap conductance for $h_{0}>0$ whose
size determines the strength of altermagnetism [Fig.\thinspace \ref{Fig2}%
(c)]. For a $d_{xy}$-wave AM and $d_{xy}$-wave SC, where the superconducting
gap nodes are intact, the surface flat bands inherent to $d_{xy}$-wave SCs
remain robust against the AM field when $h_{0}<\Delta _{0}$, see
Fig.\thinspace \ref{Fig2}(d). This leads to a ZBCP in $\sigma
_{\mathrm{SC}}^{e}(eV)$ for $h_{0}<\Delta _{0}$, but it reduces when $%
h_{0}>\Delta _{0}$ into an almost constant value within $|eV|<h_{0}-\Delta
_{0}$. Further insights are obtained by looking at the zero-bias
angle-resolved charge conductance for the $d_{xy}$-wave AM,
\begin{equation}
\bar{\sigma}_{\sigma }^{e}(eV=0)=1+\bigg|h_{0}/\Delta _{0}+\sqrt{%
h_{0}^{2}/\Delta _{0}^{2}-1}\bigg|^{-1}.  \label{dxycond}
\end{equation}%
For $h_{0}<\Delta _{0}$, the zero energy surface bound states emerge for any
value of $\theta $ within the gap and thus give rise to $\bar{\sigma}%
_{\sigma }^{e}=2$ in Eq.\thinspace (\ref{dxycond}). However, for $%
h_{0}>\Delta _{0}$, the generation of continuum states at zero energy causes
$\bar{\sigma}_{\sigma }^{e}(0)$ to decrease. From this result, we can see
that the ZBCP is influenced not only by the symmetry but also by the
strength of $d_{xy}$-wave altermagnetism. 

\begin{figure}[!t]
\includegraphics[width=88mm]{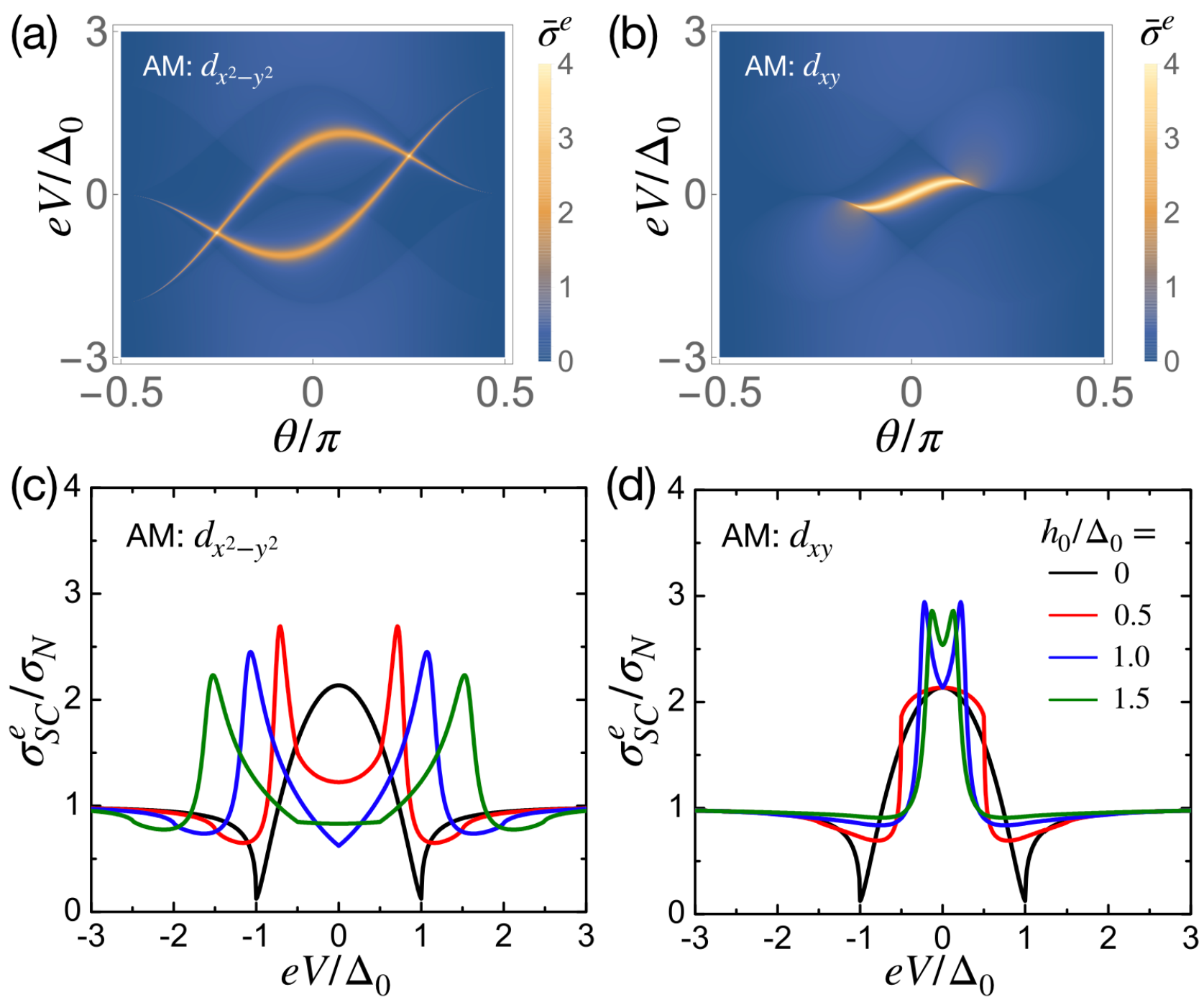}
\caption{ Conductance for junctions with chiral $p$-wave SCs. AM is $d_{x^{2}-y^{2}}$ in (a)(c) and $d_{xy}$ in (b)(d). (a)(b) angle-resolved charge conductance at $h_{0}=\Delta_{0}$. (c)(d) total conductance for distinct values of $h_{0}$. $Z=2$ for all panels.}
\label{Fig4}
\end{figure}

\emph{Surface curved bands by $d$-wave altermagnetism in chiral $p$-wave SCs.%
}--- Chiral $p$-wave pairing was once believed to describe superconductivity
in Sr$_{2}$RuO$_{4}$ with chiral edge states \cite%
{MaenoRev03,kallin2012chiral,maeno2024mystery}, and a recent study even
supports that Sr$_{2}$RuO$_{4}$ supports intrinsic AM order \cite%
{autieri2025}. Motivated by these ideas, here we consider a one-band model
with chiral $p$-wave pairing and $d$-wave altermagnetism; in the End Matter,
we also address $p$-wave pairing with $g$-wave altermagnetism. For a $%
d_{x^{2}-y^{2}}$-wave AM, we calculate the chiral SABSs by imposing $\Gamma
_{\sigma }^{+}\Gamma _{\sigma }^{-}=-e^{2i\theta }$ in Eq.\thinspace (\ref%
{condition}),
\begin{equation}
E_{\text{ABS}\nu }=\Delta _{0}\sin \theta +\nu h_{0}\cos 2\theta ,
\end{equation}%
with $\nu =\pm $. This reveals that the altermagnet field $h_{0}$ induces a
spin splitting in the chiral SABSs $E_{\text{ABS}}^{\pm }$, which appear
with a curved dispersion with $\theta $ and also revealed by angle-resolved
conductance in Fig.\thinspace \ref{Fig4}(a). The splitting
arises due $\Gamma _{\uparrow }^{+}\Gamma _{\uparrow }^{-}\neq \Gamma
_{\downarrow }^{+}\Gamma _{\downarrow }^{-}$ in a $d_{x^{2}-y^{2}}$-wave AM and there are thus two species of SABSs as shown in Fig.\thinspace \ref{Fig4}(a).
At certain angles found from $\Delta _{0}\sin \theta =\pm h_{0}\cos 2\theta $%
, however, the SABSs remain degenerate at zero energy [Fig.\thinspace \ref%
{Fig4}(a)], which affects their dispersion and makes them to acquire a
quasi-flat band profile for a range of $\theta $. These quasi-flat bands
then induce spikes in the total charge conductance whose separation for $%
h_{0}>\Delta _{0}/(2\sqrt{2})$ [Fig.\thinspace \ref%
{Fig4}(c)] is given by $\Delta _{0}^{2}/\left(
4h_{0}\right) +2h_{0}$ from $\partial E_{\text{ABS}\nu }/\partial \theta =0$
and hence determines the strength of altermagnetism.

When $d_{xy}$-wave altermagnetism is present in chiral $p$-wave SCs, we
obtain the chiral SABSs to be given by
\begin{equation}
E_{\text{ABS}\nu }=\sin \theta \sqrt{\Delta _{0}^{2}-4h_{0}^{2}\sin
^{2}\theta }\,,  \label{chiralpcond}
\end{equation}%
which shows that SABSs acquire a curved dispersion with $\theta $ but are
not split in the presence of $d_{xy}$-wave AM because here $\Gamma _{\uparrow
}^{+}\Gamma _{\uparrow }^{-}=\Gamma _{\downarrow }^{+}\Gamma _{\downarrow
}^{-}$. Moreover, Eq.\thinspace (\ref{chiralpcond}) further unveils that the
observation of such curved SABSs require $|\theta |<\arcsin
^{-1}[\Delta _{0}/\left( 2h_{0}\right) ]$, as demonstrated by the
angle-resolved charge conductance in Fig.\thinspace \ref{Fig4}%
(b). This curved dispersion of SABSs can result in a quasi-flat band at large angle and the corresponding total charge conductance then exhibits a broadened
ZBCP for $|h_{0}|<\Delta _{0}$ which then splits around $eV=0$ for $%
|h_{0}|>\Delta _{0}$, as shown in Fig.\thinspace \ref{Fig4}(d).

\begin{figure}[!b]
\includegraphics[width=88mm]{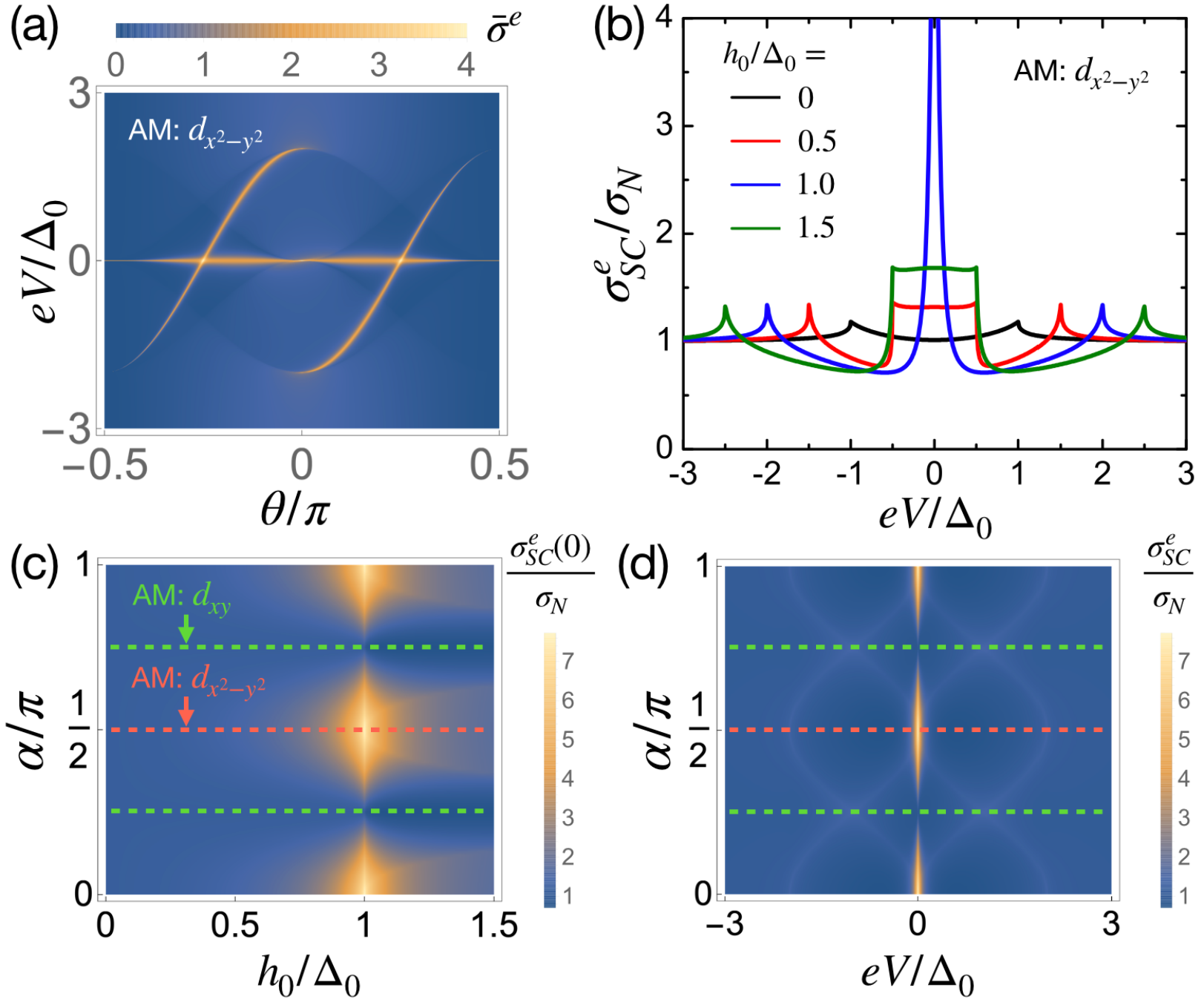}
\caption{ Conductance for junctions with chiral $d$-wave SCs with a $d$-wave AM. (a) angle-resolved charge conductance at $h_{0}=\Delta_{0}$ and (b) total conductance for distinct values of $h_{0}$ for a $d_{x^2-y^2}$-wave AM. (c) zero bias conductance vs $\alpha$ and $h_0$ where $\alpha$ denotes the crystal orientation of AM. (d) Same junction with (c) but the total conductance as a function of $eV$ and $\alpha$ at $h_0=\Delta_0$. $Z=2$ for all panels. } 
\label{Fig5}
\end{figure}

\emph{Surface flat bands by $d_{x^{2}-y^{2}}$-wave altermagnetism in chiral $%
d$-wave SCs.}--- To have a complete picture of altermagnetism inducing subgap states, here we
address it when chiral $d$-wave pairing coexists with $d$-wave AMs. In the
absence of AMs, it is well-known that chiral $d$-wave SCs host 2 chiral
SABSs \cite{BlackSchaffer_2014,Nandkishore2012,WanSheng12,Kiesel12}. Thus,
under $d_{x^{2}-y^{2}}\,$-AM, it is reasonable to expect that such SABSs are
split into 4 by $d_{x^{2}-y^{2}}\,$-AM. Actually, the obtained SABSs are
given by
\begin{equation}
E_{\text{ABS}\nu }=-\left( \Delta _{0}+\nu h_{0}\right) \text{sgn}\left(
\theta \right) \cos 2\theta ,  \label{dispcd}
\end{equation}%
where $\nu =\pm $ and $\text{sgn}(.)$ denotes the sign operation. The split
SABSs can be seen from Eq.\thinspace (\ref{dispcd}). Interestingly, we
verified that the dispersion of one SABS becomes steeper, while the other
flattens by increasing $h_{0}$ and gets completely flat at zero energy when $%
h_{0}=\Delta _{0}$ for all $k_{y}$, as shown in the angle-resolved charge
conductance in Fig.\thinspace \ref{Fig5}(a). The zero-energy flat band
further leads to a ZBCP in the total charge conductance at $h_{0}=\Delta _{0}
$ [Fig.\thinspace \ref{Fig5}(b)], while the subgap conductance becomes less
prominent for $h_{0}\neq \Delta _{0}$. For the $d_{xy}$-AM, the pair of
SABSs have no spin-splittings but acquire instead a curved dispersion due to
altermagnetism. It is noted that the emergence of ZBCP can be robust against
a certain twist between AM\ and SC. As Figs.\ref{Fig5}(c,d) show, there is a
conductance peak at $h_{0}=\Delta _{0}$ for a large range around $\alpha =\pi /2$ ($%
d_{x^{2}-y^{2}}$-AM) and disapear at $\alpha =3\pi /4$ ($d_{xy}$-AM). We thus
view it as  the resilient feature for experimental detection of the flat
SABSs in chiral $d$-wave SCs.

\emph{Summary.}---
In this work, we have demonstrated rich properties of subgap bands and surface bound states where superconductivity and altermagnetism coexist. It shows that the zero-energy states can be enhanced when the symmetry of altermagnetism and pairings are the same, leading to a zero-biased conductance peak without any formation of surface bound states. 
We verified the robustness of the flat surface Andreev bound states arising from unconventional nodal pairings, which can be seriously influenced by either the symmetry or the magnitude of altermagnetism. 
It is also found that in chiral superconductors, the chiral Andreev bound states can be split, curved, or flattened by altermagnetism. Our study reveals the importance of the anisotropic orbital symmetry as well as the lifting of spin degeneracy by altermagnetism in the formation of subgap states.   Notably, the integration of altermagnet-superconductor heterostructures is expected to induce spin-triplet pairings \cite{franze2025} and the characteristics of resulting subgap states by spin-orbit and impurity effect would be an interesting topic for future research. 
In terms of experimental test of our predictions, it
might be possible to exploit the reported altermagnets, such as KV$_{2}$Se$_{2}$ \cite{Jiang2025} and various unconventional superconductors, such as cuprates \cite{TsueiRMP}.  Moreover, recent studies have shown that
both the magnetic strength and orientation of altermagnets are electrically
controllable \cite{wang2025,Chenyiyuan25}, providing an accessible way to change altermagnetic fields. It has also been proposed that chemical doping has the potential to control the AM field \cite{Devaraj09}. 
For future applications, the subgap states are shown to modulate conductance spectra in the absence of a net magnetic field, useful for making field-free functional superconducting devices.

\emph{Acknowledgements.}---
B.\ L.\  thanks F. M. Qu for useful discussions related to this work and acknowledges financial support from the National Natural Science Foundation of China (project 12474049) and Beijing National Laboratory for Condensed Matter Physics (2025BNLCMPKF011).   
P.M. and P.B. acknowledge support by the Spanish CM ``Talento Program'' project No.~2019-T1/IND-14088 and No.~2023-5A/IND-28927, the Agencia Estatal de Investigaci\'on project No.~PID2020-117992GA-I00, No.~PID2024-157821NB-I00 and No.~CNS2022-135950 and through the ``María de Maeztu'' Programme for Units of Excellence in R\&D (CEX2023-001316-M).  
J.\ C.\ acknowledges financial support from the Swedish Research Council  (Vetenskapsr\aa det Grant No.~2021-04121) and the Carl Trygger’s Foundation (Grant No. 22: 2093). 
Y.\ T. \ acknowledges financial support from JSPS with Grants-in-Aid for Scientific Research (KAKENHI Grants Nos. 23K17668, 24K00583, 24K00556, 24K00578, 25H00609 and 25H00613). 
Y.\ F.\ acknowledges financial support from the Sumitomo Foundation and numerical calculation support from Okayama University.

\bibliography{biblio}

\end{document}

\title{Supplementary material for ``Engineering subgap states in
superconductors by the symmetry of altermagnetism''}
\newcommand{\tianjin}{Department of Physics, Tianjin University, Tianjin 300354, China}
	
\newcommand{\nagoya}{Department of Applied Physics, Nagoya University, Nagoya 464-8603, Japan}
	
\newcommand{\Okayama}{Faculty of Environmental Life, Natural Science and Technology, Okayama University, 700-8530 Okayama, Japan}
	
\newcommand{\Uppsala}{Department of Physics and Astronomy, Uppsala University, Box 516, S-751 20 Uppsala, Sweden}
	
\newcommand{\uam}{Department of Theoretical Condensed Matter Physics, Universidad Aut\'onoma de Madrid, 28049 Madrid, Spain}
\newcommand{\ifimac}{Condensed Matter Physics Center (IFIMAC), Universidad Aut\'onoma de Madrid, 28049 Madrid, Spain}
\newcommand{\inc}{Instituto Nicol\'as Cabrera, Universidad Aut\'onoma de Madrid, 28049 Madrid, Spain}

\author{Bo Lu}
\affiliation{\tianjin}

\author{Phillip Mercebach}
\affiliation{\uam}
\affiliation{\ifimac}

\author{Pablo Burset}
\affiliation{\uam}
\affiliation{\ifimac}
\affiliation{\inc}

\author{Keiji Yada}
\affiliation{\nagoya}
	
\author{Jorge Cayao}
\affiliation{\Uppsala}
	
\author{Yukio Tanaka}
\affiliation{\nagoya}
	
\author{Yuri Fukaya}
\affiliation{\Okayama}

\maketitle

%\widetext

%%%%%%%%%% Merge with supplemental materials %%%%%%%%%%
%\widetext
%%%%%%%%%% Merge with supplemental materials %%%%%%%%%%
%%%%%%%%%% Prefix a "S" to all equations, figures, tables and reset the counter %%%%%%%%%%
\setcounter{equation}{0} \setcounter{figure}{0} \setcounter{table}{0}  %
\setcounter{page}{1} \makeatletter
\renewcommand{\theequation}{S\arabic{equation}} \renewcommand{%
\thefigure}{S\arabic{figure}} \renewcommand{\bibnumfmt}[1]{[S#1]} %
\renewcommand{\citenumfont}[1]{S#1}

\noindent \textbf{S1. Details of the conductance in the tight-binding model}\newline

In this section, we present the formulation to calculate the tunneling
conductance in the tight-binding model.

\begin{figure}[tbp]
\centering
\includegraphics[width=8.5cm]{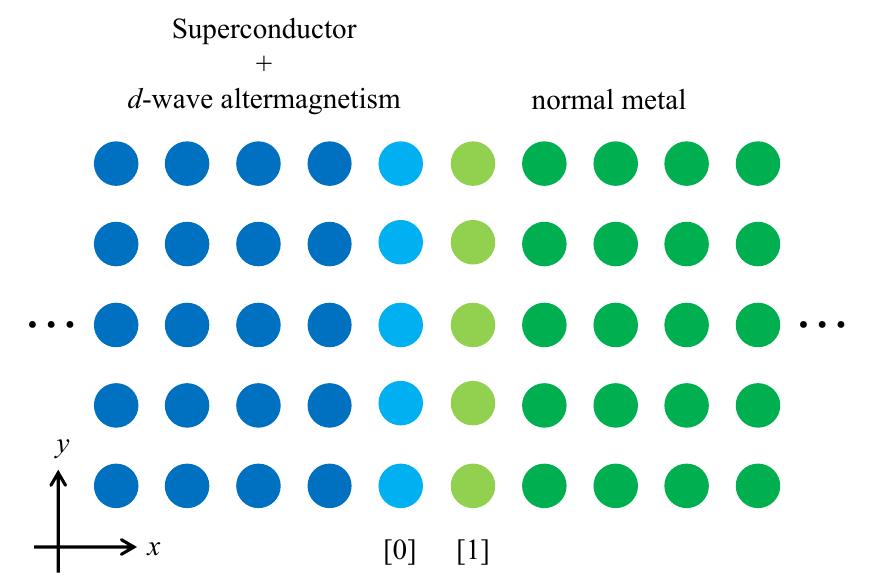}
\caption{Image of superconductor with $d$-wave altermagnetism/normal metal
junctions in the tight-binding (lattice) model. 0 and 1 correspond to the
indices in the local and nonlocal Green's functions. We assume the periodic
boundary condition along the $y$-direction and the semi-infinite
superconductors and normal metal along the $x$-direction.}
\label{fig:image_TBM}
\end{figure}

\begin{figure}[t!]
\centering
\includegraphics[width=17.5cm]{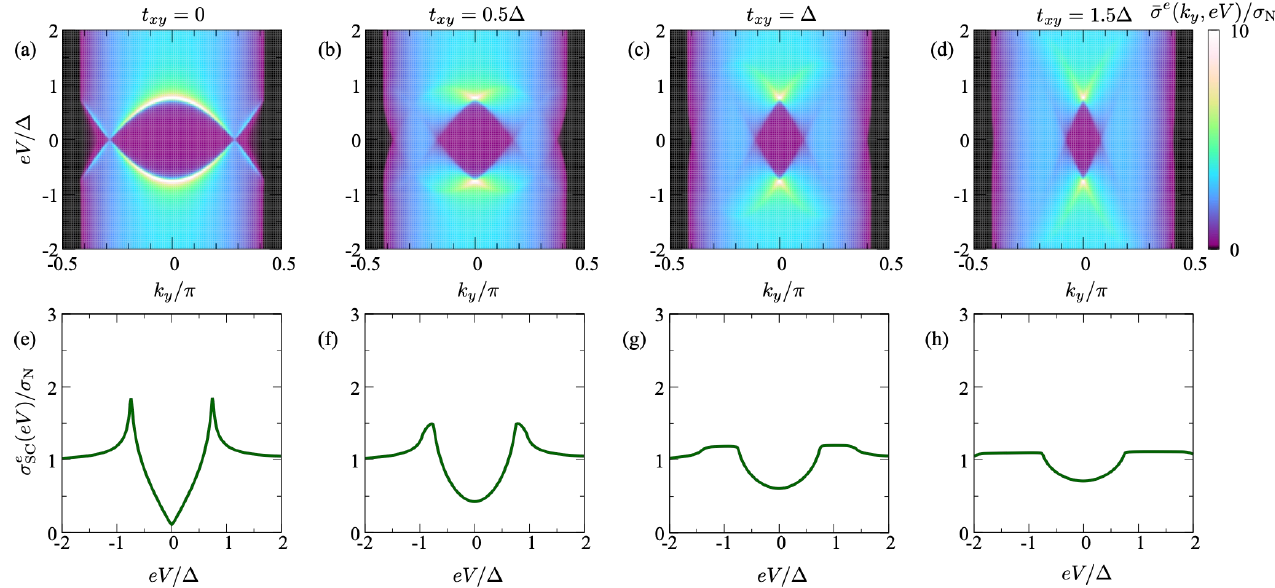}
\caption{(a)(b)(c)(d) Momentum-resolved tunneling conductance in
spin-singlet $d_{x^2-y^2}$-wave superconductor with $d_{xy}$-wave
altermagnetism/normal metal junctions at (a) $t_{xy}=0$, (b) $%
t_{xy}=0.5\Delta$, (c) $t_{xy}=\Delta$ and (d) $t_{xy}=1.5\Delta$. $\protect%
\sigma_\mathrm{N}$ means the tunneling conductance at zero energy in the
normal state. (e)(f)(g)(h)(i) Tunneling conductance in spin-singlet $%
d_{x^2-y^2}$-wave superconductor with $d_{xy}$-wave altermagnetism/normal
metal junctions at (a) $t_{xy}=0$, (b) $t_{xy}=0.5\Delta$, (c) $t_{xy}=\Delta
$ and (d) $t_{xy}=1.5\Delta$. Parameters: $\protect\mu=-2.5t$, $t_{1}=t$, $%
\Delta=0.01t$, $U_b=5t$, and $\protect\delta=0.01\Delta$.}
\label{fig:dx2_y2_wave_dxy_AM}
\end{figure}
\begin{figure}[t!]
\centering
\includegraphics[width=17.5cm]{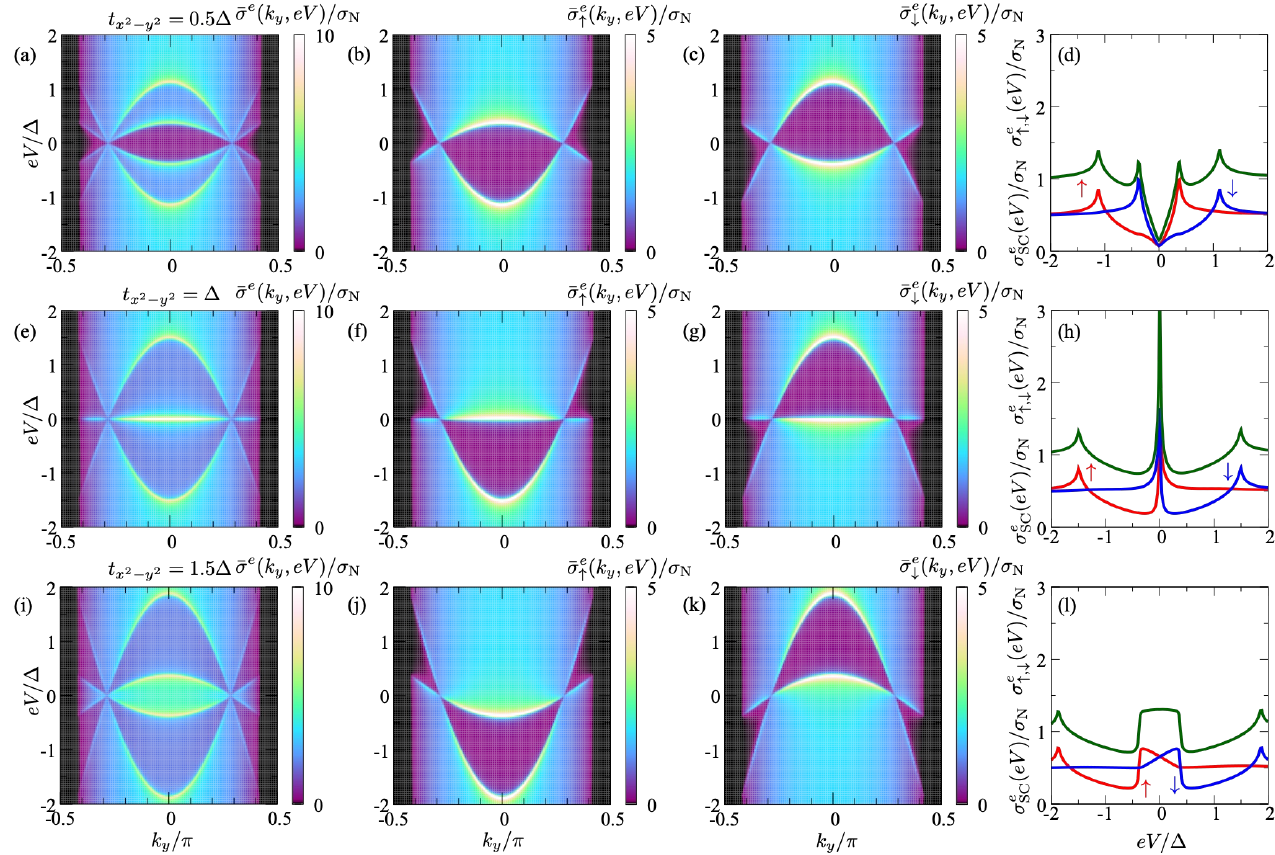}
\caption{(a)(e)(i) Momentum-resolved tunneling conductance in spin-singlet $%
d_{x^2-y^2}$-wave superconductor with $d_{x^2-y^2}$-wave
altermagnetism/normal metal junctions at (a) $t_{x^2-y^2}=0.5\Delta$, (e) $%
t_{x^2-y^2}=\Delta$, and (i) $t_{x^2-y^2}=1.5\Delta$. $\protect\sigma_%
\mathrm{N}$ means the tunneling conductance at zero energy in the normal
state. (b)(c)(f)(g)(j)(k) Momentum-resolved tunneling conductance with
(b)(f)(j) up-spin and (c)(g)(k) down-spin in spin-singlet $d_{x^2-y^2}$-wave
superconductor with $d_{x^2-y^2}$-wave altermagnetism/normal metal junctions
at (b)(c) $t_{x^2-y^2}=0.5\Delta$, (f)(g) $t_{x^2-y^2}=\Delta$, and (j)(k) $%
t_{x^2-y^2}=1.5\Delta$. (d)(h)(l) Tunneling conductance $\protect\sigma^{e}_%
\mathrm{SC}(eV)$ and spin-resolved ones $\protect\sigma^{e}_{\uparrow,%
\downarrow}(eV)$ in spin-singlet $d_{x^2-y^2}$-wave superconductor with $%
d_{x^2-y^2}$-wave altermagnetism/normal metal junctions at (b)(c) $%
t_{x^2-y^2}=0.5\Delta$, (f)(g) $t_{x^2-y^2}=\Delta$, and (j)(k) $%
t_{x^2-y^2}=1.5\Delta$. In (d)(h)(l), the green line indicates the tunneling
conductance, and the red and blue ones mean the tunneling conductance with
up and down-spin, respectively. Parameters: $\protect\mu=-2.5t$, $t_{1}=t$, $%
\Delta=0.01t$, $U_b=5t$, and $\protect\delta=0.01\Delta$.}
\label{fig:dx2_y2_wave_dx2_y2_AM}
\end{figure}
\begin{figure}[t!]
\centering
\includegraphics[width=17.5cm]{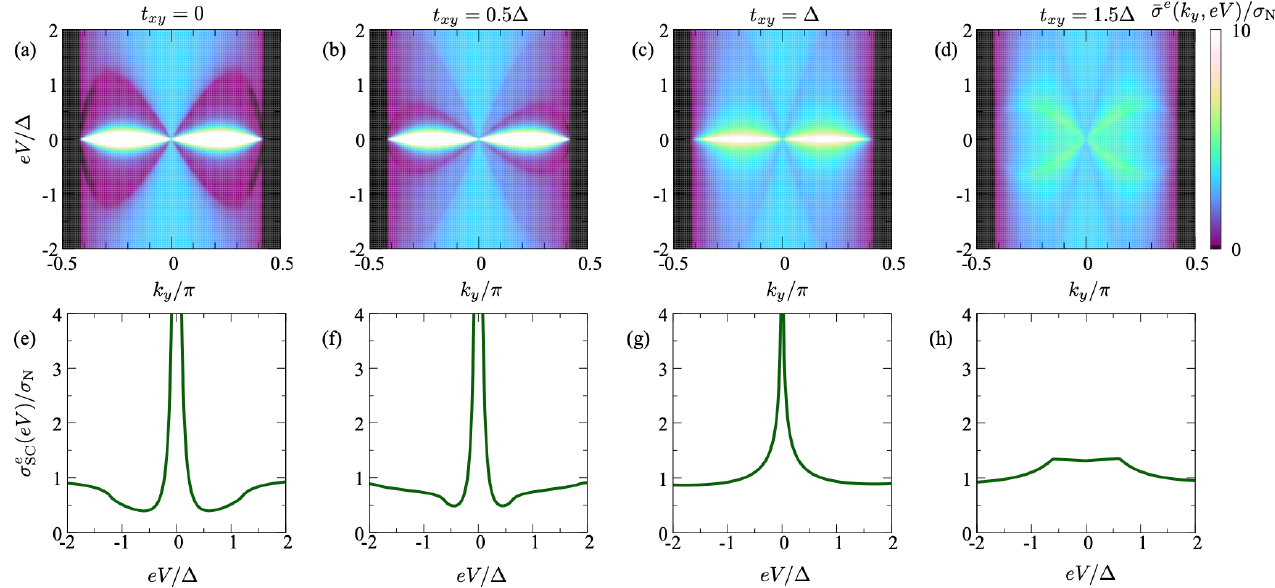}
\caption{(a)(b)(c)(d) Momentum-resolved tunneling conductance in
spin-singlet $d_{xy}$-wave superconductors with $d_{xy}$-wave
altermagnetism/normal metal junctions at (a) $t_{xy}=0$, (b) $%
t_{xy}=0.5\Delta$, (c) $t_{xy}=\Delta$ and (d) $t_{xy}=1.5\Delta$. $\protect%
\sigma_\mathrm{N}$ means the tunneling conductance at zero energy in the
normal state. (e)(f)(g)(h)(i) Tunneling conductance in spin-singlet $d_{xy}$%
-wave superconductor with $d_{xy}$-wave altermagnetism/normal metal
junctions at (a) $t_{xy}=0$, (b) $t_{xy}=0.5\Delta$, (c) $t_{xy}=\Delta$ and
(d) $t_{xy}=1.5\Delta$. Parameters: $\protect\mu=-2.5t$, $t_{1}=t$, $%
\Delta=0.01t$, $U_b=5t$, and $\protect\delta=0.01\Delta$.}
\label{fig:dxy_wave_dxy_AM}
\end{figure}
\begin{figure}[t!]
\centering
\includegraphics[width=17.5cm]{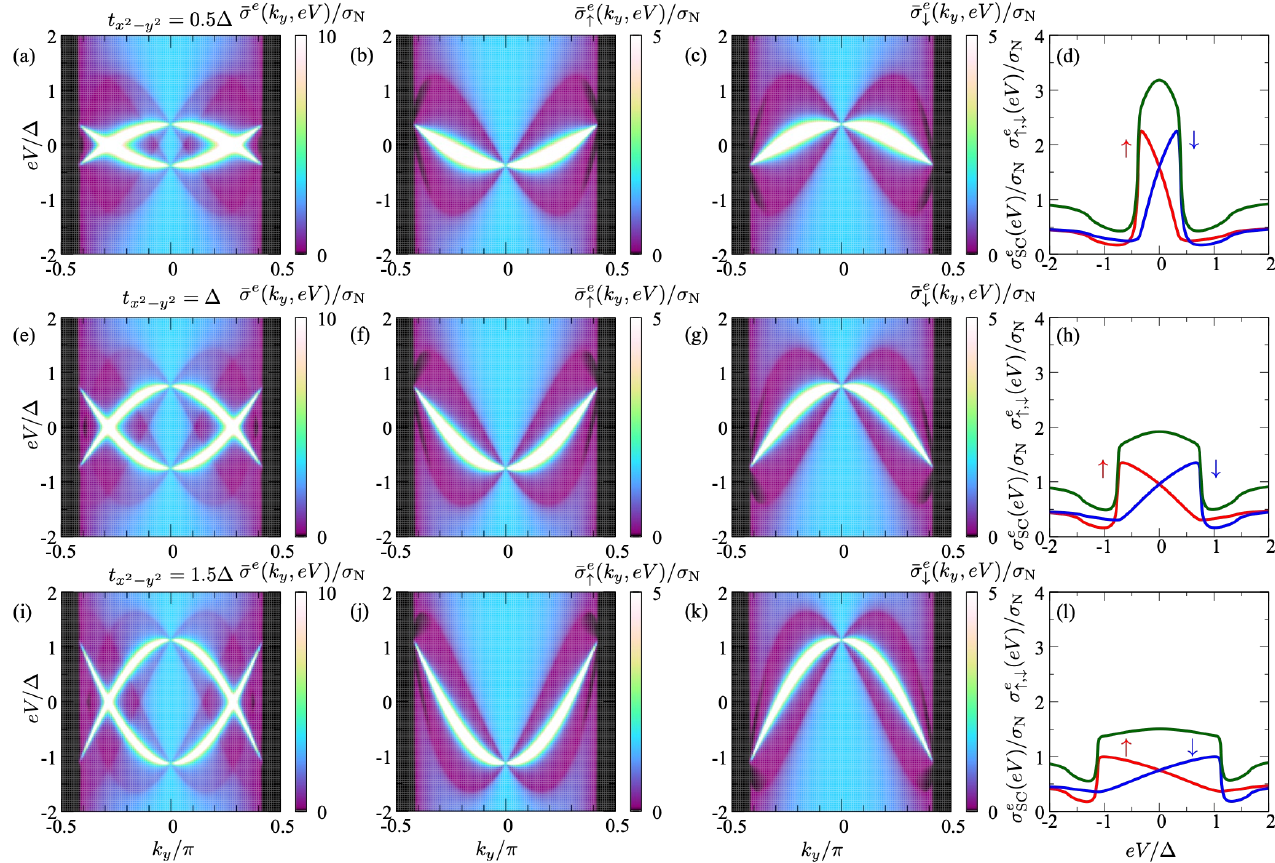}
\caption{(a)(e)(i) Momentum-resolved tunneling conductance in spin-singlet $%
d_{xy}$-wave superconductor with $d_{x^2-y^2}$-wave altermagnetism/normal
metal junctions at (a) $t_{x2-y^2}=0.5\Delta$, (e) $t_{x2-y^2}=\Delta$, and
(i) $t_{x2-y^2}=1.5\Delta$. $\protect\sigma_\mathrm{N}$ means the tunneling
conductance at zero energy in the normal state. (b)(c)(f)(g)(j)(k)
Momentum-resolved tunneling conductance with (b)(f)(j) up-spin and (c)(g)(k)
down-spin in spin-singlet $d_{xy}$-wave superconductor with $d_{x^2-y^2}$%
-wave altermagnetism/normal metal junctions at (b)(c) $t_{x2-y^2}=0.5\Delta$%
, (f)(g) $t_{x2-y^2}=\Delta$, and (j)(k) $t_{x2-y^2}=1.5\Delta$. (d)(h)(l)
Tunneling conductance $\protect\sigma^{e}_\mathrm{SC}(eV)$ and spin-resolved
ones $\protect\sigma^{e}_{\uparrow,\downarrow}(eV)$ in spin-singlet $d_{xy}$%
-wave superconductor with $d_{x^2-y^2}$-wave altermagnetism/normal metal
junctions at (b)(c) $t_{x2-y^2}=0.5\Delta$, (f)(g) $t_{x2-y^2}=\Delta$, and
(j)(k) $t_{x2-y^2}=1.5\Delta$. In (d)(h)(l), the green line indicates the
tunneling conductance, and the red and blue ones mean the tunneling
conductance with up and down-spin, respectively. Parameters: $\protect\mu%
=-2.5t$, $t_{1}=t$, $\Delta=0.01t$, $U_b=5t$, and $\protect\delta=0.01\Delta$%
.}
\label{fig:dxy_wave_dx2_y2_AM}
\end{figure}
\begin{figure}[t!]
\centering
\includegraphics[width=17.5cm]{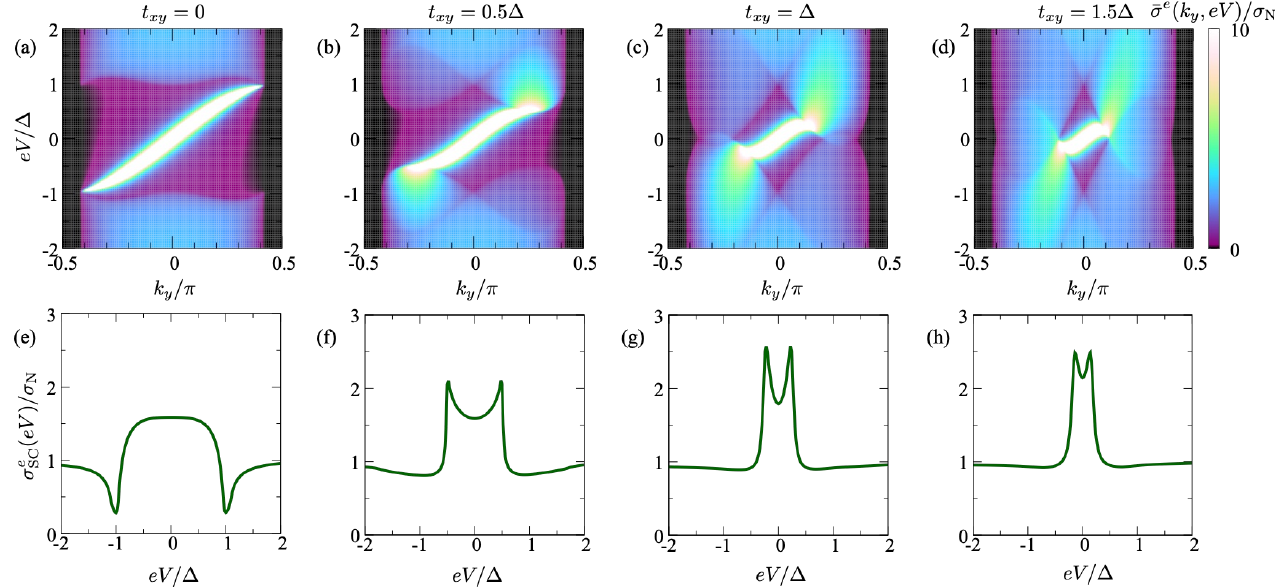}
\caption{(a)(b)(c)(d) Momentum-resolved tunneling conductance in
spin-triplet chiral $p$-wave superconductor with $d_{xy}$-wave
altermagnetism/normal metal junctions at (a) $t_{xy}=0$, (b) $%
t_{xy}=0.5\Delta$, (c) $t_{xy}=\Delta$ and (d) $t_{xy}=1.5\Delta$. $\protect%
\sigma_\mathrm{N}$ means the tunneling conductance at zero energy in the
normal state. (e)(f)(g)(h)(i) Tunneling conductance in spin-triplet chiral $p
$-wave superconductor with $d_{xy}$-wave altermagnetism/normal metal
junctions at (a) $t_{xy}=0$, (b) $t_{xy}=0.5\Delta$, (c) $t_{xy}=\Delta$ and
(d) $t_{xy}=1.5\Delta$. Parameters: $\protect\mu=-2.5t$, $t_{1}=t$, $%
\Delta=0.01t$, $U_b=5t$, and $\protect\delta=0.01\Delta$.}
\label{fig:chiral_p_wave_dxy_AM}
\end{figure}
\begin{figure}[t!]
\centering
\includegraphics[width=17.5cm]{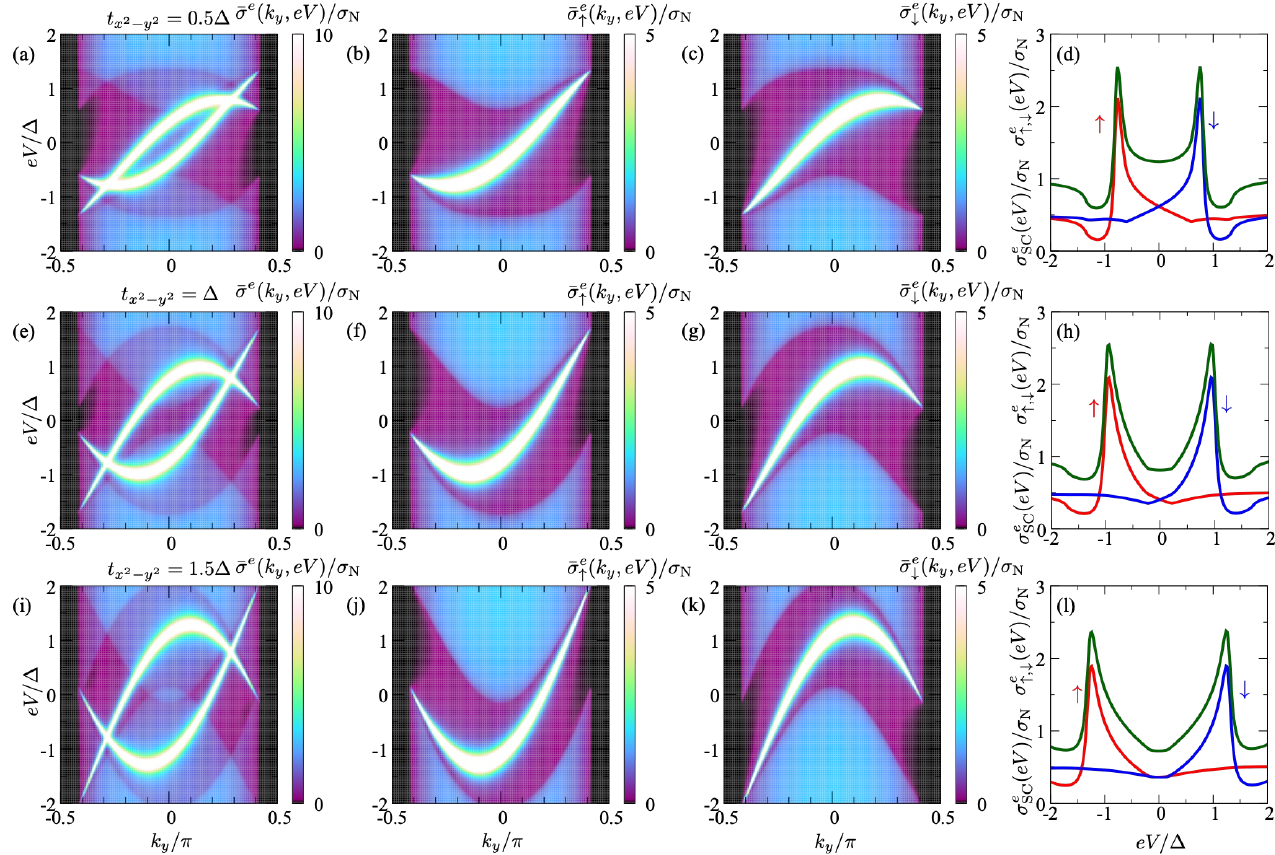}
\caption{(a)(e)(i) Momentum-resolved tunneling conductance in spin-triplet
chiral $p$-wave superconductor with $d_{x^2-y^2}$-wave altermagnetism/normal
metal junctions at (a) $t_{x2-y^2}=0.5\Delta$, (e) $t_{x2-y^2}=\Delta$, and
(i) $t_{x2-y^2}=1.5\Delta$. $\protect\sigma_\mathrm{N}$ means the tunneling
conductance at zero energy in the normal state. (b)(c)(f)(g)(j)(k)
Momentum-resolved tunneling conductance with (b)(f)(j) up-spin and (c)(g)(k)
down-spin in spin-triplet chiral $p$-wave superconductor with $d_{x^2-y^2}$%
-wave altermagnetism/normal metal junctions at (b)(c) $t_{x2-y^2}=0.5\Delta$%
, (f)(g) $t_{x2-y^2}=\Delta$, and (j)(k) $t_{x2-y^2}=1.5\Delta$. (d)(h)(l)
Tunneling conductance $\protect\sigma^{e}_\mathrm{SC}(eV)$ and spin-resolved
ones $\protect\sigma^{e}_{\uparrow,\downarrow}(eV)$ in spin-triplet chiral $p
$-wave superconductor with $d_{x^2-y^2}$-wave altermagnetism/normal metal
junctions at (b)(c) $t_{x2-y^2}=0.5\Delta$, (f)(g) $t_{x2-y^2}=\Delta$, and
(j)(k) $t_{x2-y^2}=1.5\Delta$. In (d)(h)(l), the green line indicates the
tunneling conductance, and the red and blue ones mean the tunneling
conductance with up and down-spin, respectively. Parameters: $\protect\mu%
=-2.5t$, $t_{1}=t$, $\Delta=0.01t$, $U_b=5t$, and $\protect\delta=0.01\Delta$%
.}
\label{fig:chiral_p_wave_dx2_y2_AM}
\end{figure}
\begin{figure}[t!]
\centering
\includegraphics[width=17.5cm]{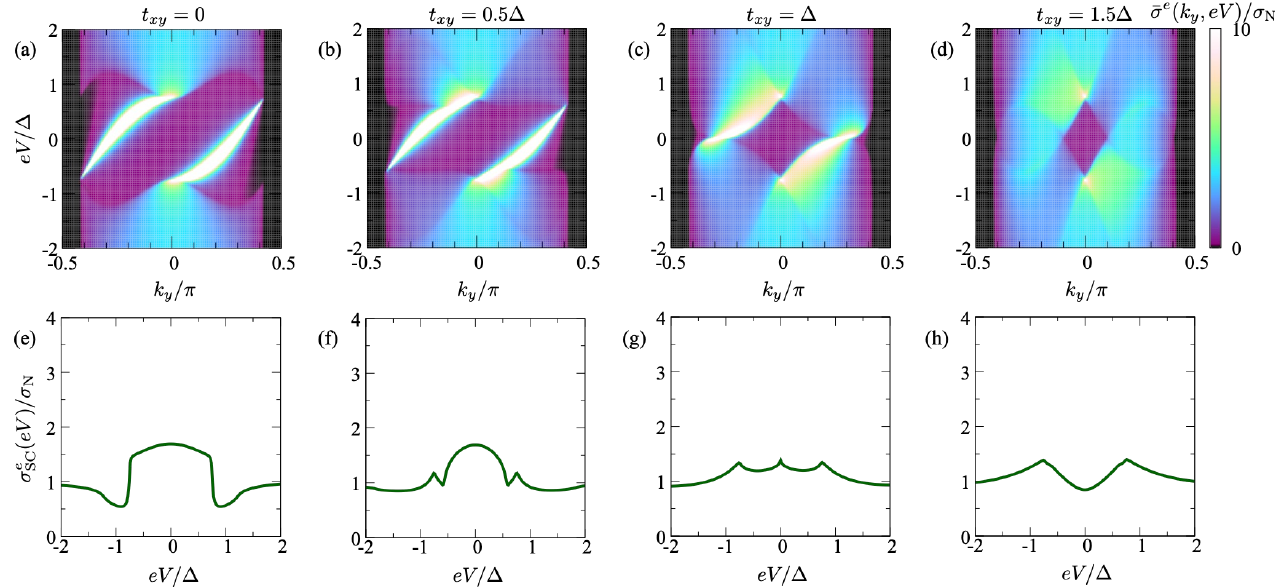}
\caption{(a)(b)(c)(d) Momentum-resolved tunneling conductance in
spin-singlet chiral $d$-wave superconductor with $d_{xy}$-wave
altermagnetism/normal metal junctions at (a) $t_{xy}=0$, (b) $%
t_{xy}=0.5\Delta$, (c) $t_{xy}=\Delta$ and (d) $t_{xy}=1.5\Delta$. $\protect%
\sigma_\mathrm{N}$ means the tunneling conductance at zero energy in the
normal state. (e)(f)(g)(h)(i) Tunneling conductance in spin-singlet chiral $d
$-wave superconductor with $d_{xy}$-wave altermagnetism/normal metal
junctions at (a) $t_{xy}=0$, (b) $t_{xy}=0.5\Delta$, (c) $t_{xy}=\Delta$ and
(d) $t_{xy}=1.5\Delta$. Parameters: $\protect\mu=-2.5t$, $t_{1}=t$, $%
\Delta=0.01t$, $U_b=5t$, and $\protect\delta=0.01\Delta$.}
\label{fig:chiral_d_wave_dxy_AM}
\end{figure}
\begin{figure}[t!]
\centering
\includegraphics[width=17.5cm]{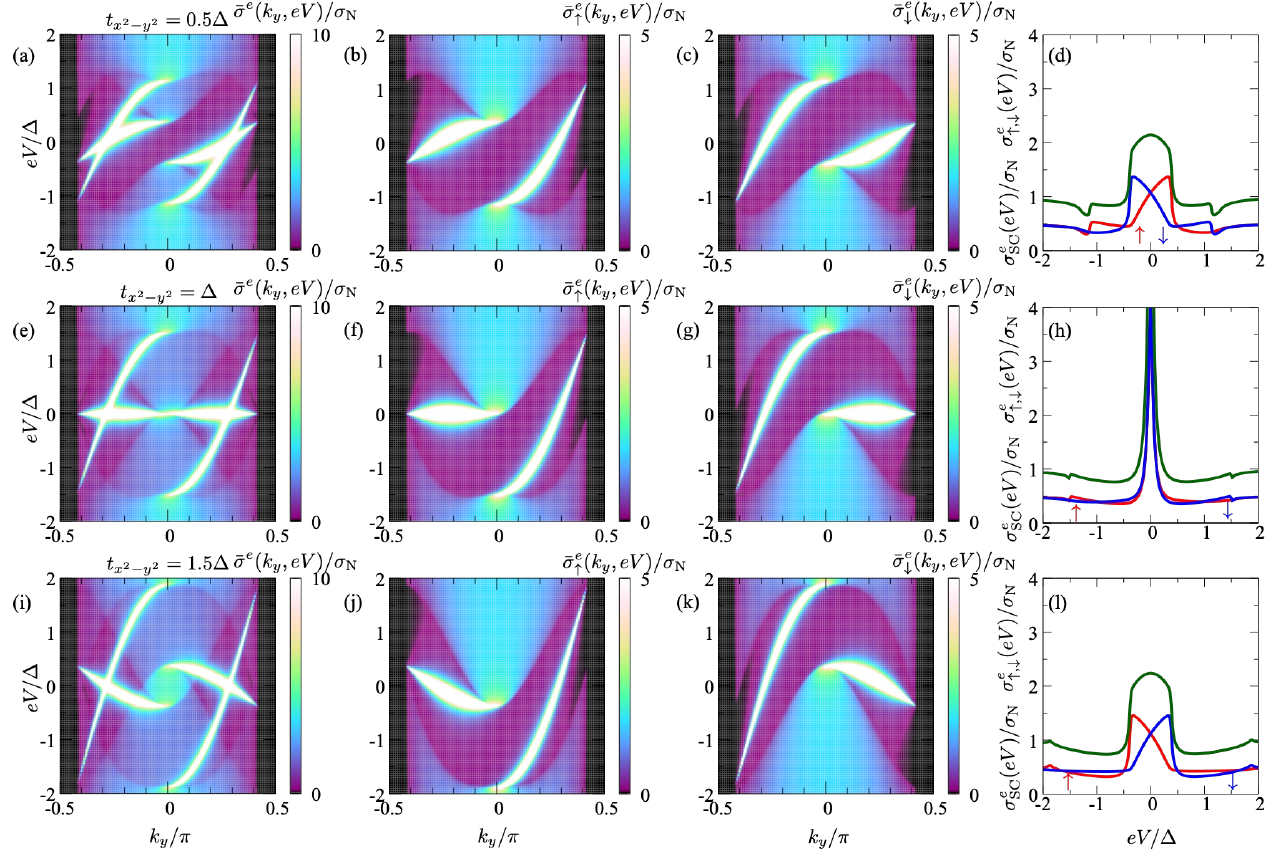}
\caption{(a)(e)(i) Momentum-resolved tunneling conductance in spin-singlet
chiral $d$-wave superconductor with $d_{x^2-y^2}$-wave altermagnetism/normal
metal junctions at (a) $t_{x2-y^2}=0.5\Delta$, (e) $t_{x2-y^2}=\Delta$, and
(i) $t_{x2-y^2}=1.5\Delta$. $\protect\sigma_\mathrm{N}$ means the tunneling
conductance at zero energy in the normal state. (b)(c)(f)(g)(j)(k)
Momentum-resolved tunneling conductance with (b)(f)(j) up-spin and (c)(g)(k)
down-spin in spin-singlet chiral $d$-wave superconductor with $d_{x^2-y^2}$%
-wave altermagnetism/normal metal junctions at (b)(c) $t_{x2-y^2}=0.5\Delta$%
, (f)(g) $t_{x2-y^2}=\Delta$, and (j)(k) $t_{x2-y^2}=1.5\Delta$. (d)(h)(l)
Tunneling conductance $\protect\sigma^{e}_\mathrm{SC}(eV)$ and spin-resolved
ones $\protect\sigma^{e}_{\uparrow,\downarrow}(eV)$ in spin-singlet chiral $d
$-wave superconductor with $d_{x^2-y^2}$-wave altermagnetism/normal metal
junctions at (b)(c) $t_{x2-y^2}=0.5\Delta$, (f)(g) $t_{x2-y^2}=\Delta$, and
(j)(k) $t_{x2-y^2}=1.5\Delta$. In (d)(h)(l), the green line indicates the
tunneling conductance, and the red and blue ones mean the tunneling
conductance with up and down-spin, respectively. Parameters: $\protect\mu%
=-2.5t$, $t_{1}=t$, $\Delta=0.01t$, $U_b=5t$, and $\protect\delta=0.01\Delta$%
.}
\label{fig:chiral_d_wave_dx2_y2_AM}
\end{figure}
\begin{figure}[t!]
\centering
\includegraphics[width=17.5cm]{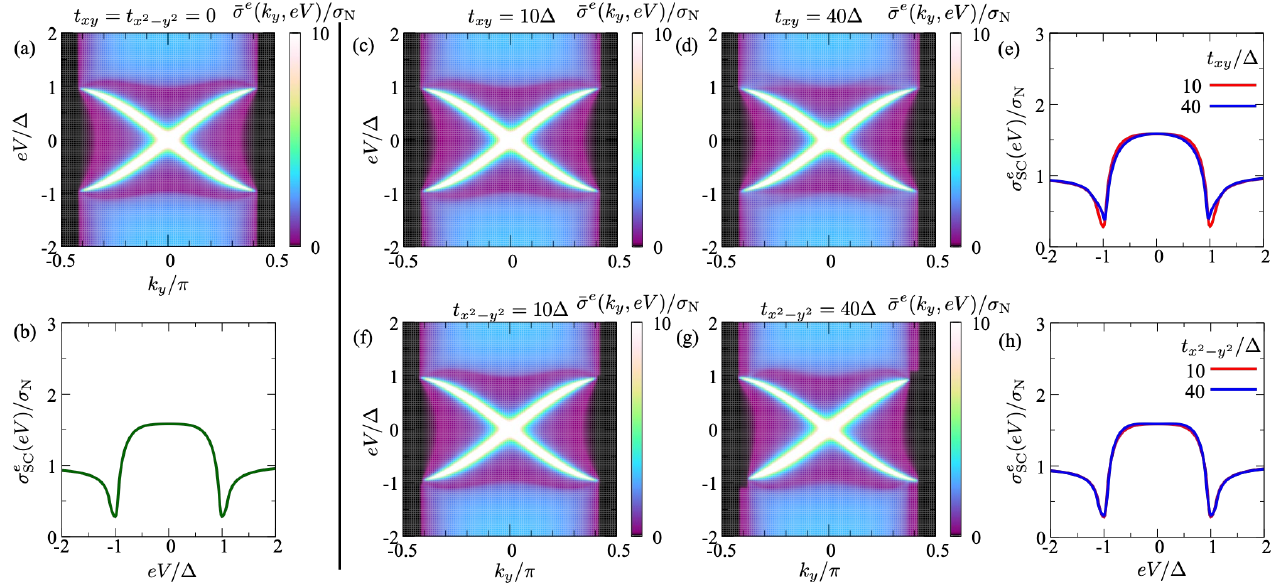}
\caption{(a) Momentum-resolved tunneling conductance in spin-triplet helical
$p$-wave superconductor/normal metal junctions. (b) Tunneling conductance
normalized by that at zero energy in the normal state $\protect\sigma_%
\mathrm{N}$ in spin-triplet helical $p$-wave superconductor/normal metal
junctions. (c)(d)(f)(h) Momentum-resolved tunneling conductance in
spin-triplet helical $p$-wave superconductor with $d$-wave
altermagnetism/normal metal junctions at (c) $t_{xy}=10\Delta$, (d) $%
t_{xy}=40\Delta$, (f) $t_{x^2-y^2}=10\Delta$ and (g) $t_{x2-y^2}=40\Delta$.
(e)(h) Tunneling conductance $\protect\sigma^{e}_\mathrm{SC}(eV)$ normalized
by $\protect\sigma_\mathrm{N}$ in spin-triplet helical $p$-wave
superconductor with $d$-wave altermagnetism/normal metal junctions at (e) $%
t_{xy}=10\Delta$ (red) and $t_{xy}=40\Delta$ (blue lines), and (h) $%
t_{x2-y^2}=10\Delta$ (red) and $t_{x2-y^2}=40\Delta$ (blue lines).
Parameters: $\protect\mu=-2.5t$, $t_{1}=t$, $\Delta=0.01t$, $U_b=5t$, and $%
\protect\delta=0.01\Delta$.}
\label{fig:helical_p_wave_d_AM}
\end{figure}

We assume the single-orbital model, and then the Hamiltonian in the normal
state is given by
\begin{align}
\hat{H}_{0}(\bm{k})&=\varepsilon(\bm{k})\hat{\sigma}_{0}, \\
\varepsilon(\bm{k})&=-\mu-2t_1\cos{k_x}-2t_1\cos{k_y},
\end{align}
where $\hat{\sigma}_{j=0,1,2,3}$ is the Pauli matrices in spin space, $%
\mu=-2.5t$ is the chemical potential, $t_{1}=t$ is the hopping integral.  In
superconductors, we assume the coexistence of the $d$-wave altermagnetism:
\begin{align}
\hat{H}(\bm{k})=\hat{H}_{0}(\bm{k})+\hat{H}_\mathrm{AM}(\bm{k}),
\end{align}
with
\begin{align}
\hat{H}_\mathrm{AM}(\bm{k})&=[2t_{xy}\sin{k_x}\sin{k_y}+t_{x^2-y^2}(\cos{k_x}%
-\cos{k_y})]\hat{\sigma}_{3},
\end{align}
with the strength of the $d_{xy}$ and $d_{x^2-y^2}$-wave altermagnetic order
$t_{xy}$ and $t_{x^2-y^2}$.  The Bogoliubov--de Gennes Hamiltonian in
superconductors is described by
\begin{align}
\hat{H}_\mathrm{BdG}(\bm{k})=
\begin{pmatrix}
\hat{H}(\bm{k}) & \hat{\Delta}(\bm{k}) \\
\hat{\Delta}^{\dagger}(\bm{k}) & -\hat{H}^{*}(-\bm{k})%
\end{pmatrix}%
,
\end{align}
and in the normal metal:
\begin{align}
\hat{H}_\mathrm{BdG}(\bm{k})=
\begin{pmatrix}
\hat{H}_{0}(\bm{k}) & 0 \\
0 & -\hat{H}^{*}_{0}(-\bm{k})%
\end{pmatrix}%
.
\end{align}
Here, $\hat{\Delta}(\bm{k})$ is the pair potential given by
\begin{align}
\hat{\Delta}(\bm{k})=[\psi(\bm{k})+\bm{d}(\bm{k})\cdot\hat{\bm{\sigma}} ]i%
\hat{\sigma}_{2},
\end{align}
with the spin-singlet $\psi(\bm{k})$ and spin-triplet order parameters $%
\bm{d}(\bm{k})=[d_x(\bm{k}),d_y(\bm{k}),d_z(\bm{k})]$.  In the spin-singlet
state, we assume
\begin{align}
\psi(\bm{k})=\Delta,
\end{align}
for $s$-wave,
\begin{align}
\psi(\bm{k})=\Delta(\cos{k_x}-\cos{k_y}),
\end{align}
for $d_{x^2-y^2}$-wave,
\begin{align}
\psi(\bm{k})=2\Delta\sin{k_x}\sin{k_y},
\end{align}
for $d_{xy}$-wave, and
\begin{align}
\psi(\bm{k})=\Delta[(\cos{k_x}-\cos{k_y})+2i\sin{k_x}\sin{k_y}],
\end{align}
for two-dimensional (2D) chiral $d$-wave pairings.  Likewise, in the
spin-triplet state, we consider
\begin{align}
d_x(\bm{k})&=\Delta\sin{k_y},  \notag \\
d_y(\bm{k})&=-\Delta\sin{k_x}, \\
d_z(\bm{k})&=0,  \notag
\end{align}
for the helical $p$-wave and
\begin{align}
d_x(\bm{k})&=0,  \notag \\
d_y(\bm{k})&=0, \\
d_z(\bm{k})&=\Delta(\sin{k_x}+i\sin{k_y}),  \notag
\end{align}
for the chiral $p$-wave pairings, with the energy gap amplitude $\Delta=0.01t
$.

In the superconductor with $d$-wave altermagnetism/normal metal junctions
shown in Fig.~\ref{fig:image_TBM}, we consider the Hamiltonian:
\begin{align}
\hat{H}=\hat{H}_\mathrm{SC+AM}+\hat{H}_\mathrm{J}+\hat{H}_\mathrm{NM}.
\end{align}
In the part of superconductors with altermagnetism $\hat{H}_\mathrm{SC+AM}$,
the local $\tilde{u}_\mathrm{SC}(k_y)$ and nonlocal terms $\tilde{t}_\mathrm{%
SC}(k_y)$ are given by
\begin{align}
\tilde{u}_\mathrm{SC}(k_y)&=
\begin{pmatrix}
\hat{u}_{0}(k_y)+\hat{u}_\mathrm{AM}(k_y) & \hat{u}_\mathrm{\Delta}(k_y) \\
\hat{u}^{\dagger}_{\Delta}(k_y) & -\hat{u}^{*}_{0}(-k_y)-\hat{u}^{*}_\mathrm{%
AM}(-k_y)%
\end{pmatrix}%
, \\
\tilde{t}_\mathrm{SC}(k_y)&=
\begin{pmatrix}
\hat{t}_{0}(k_y)+\hat{t}_\mathrm{AM}(k_y) & \hat{t}_\mathrm{\Delta}(k_y) \\
\bar{t}_{\Delta}(k_y) & -\hat{t}^{*}_{0}(-k_y)-\hat{t}^{*}_\mathrm{AM}(-k_y)%
\end{pmatrix}%
,
\end{align}
with
\begin{align}
\hat{u}_{0}(k_y)&=[-\mu-2t_{1}\cos{k_y}]\hat{\sigma}_{0}, \\
\hat{u}_\mathrm{AM}(k_y)&=-t_{x^2-y^2}\cos{k_y}\hat{\sigma}_{3},
\end{align}
\begin{align}
\hat{t}_{0}&=-t_1\hat{\sigma}_{0}, \\
\hat{t}_\mathrm{AM}(k_y)&=\left[-it_{xy}\sin{k_y}+\frac{t_{x^2-y^2}}{2}%
\right]\hat{\sigma}_{3}.
\end{align}
In the pair potential, we obtain
\begin{align}
\hat{u}_{\Delta}&=\Delta i\hat{\sigma}_{2}, \\
\hat{t}_{\Delta}&=\bar{t}_{\Delta}=0,
\end{align}
for spin-singlet $s$-wave,
\begin{align}
\hat{u}_{\Delta}(k_y)&=-\Delta\cos{k_y}i\hat{\sigma}_{2}, \\
\hat{t}_{\Delta}&=\frac{\Delta}{2}i\hat{\sigma}_{2}, \\
\bar{t}_{\Delta}&=-\frac{\Delta}{2}i\hat{\sigma}_{2},
\end{align}
for spin-singlet $d_{x^2-y^2}$-wave,
\begin{align}
\hat{u}_{\Delta}&=0, \\
\hat{t}_{\Delta}&=\bar{t}_{\Delta}=-i\Delta\sin{k_y}i\hat{\sigma}_{2},
\end{align}
for spin-singlet $d_{xy}$-wave,
\begin{align}
\hat{u}_{\Delta}&=-\Delta\cos{k_y}i\hat{\sigma}_{2}, \\
\hat{t}_{\Delta}&=\left[\frac{\Delta}{2}+\Delta\sin{k_y}\right]i\hat{\sigma}%
_{2}, \\
\bar{t}_{\Delta}&=\left[-\frac{\Delta}{2}+\Delta\sin{k_y}\right]i\hat{\sigma}%
_{2},
\end{align}
for 2D chiral $d$-wave,
\begin{align}
\hat{u}_{\Delta}&=
\begin{pmatrix}
-\Delta\sin{k_y} & 0 \\
0 & \Delta\sin{k_y}%
\end{pmatrix}%
, \\
\hat{t}_{\Delta}&=
\begin{pmatrix}
-\frac{\Delta}{2} & 0 \\
0 & -\frac{\Delta}{2}%
\end{pmatrix}%
, \\
\bar{t}_{\Delta}&=
\begin{pmatrix}
\frac{\Delta}{2} & 0 \\
0 & \frac{\Delta}{2}%
\end{pmatrix}%
,
\end{align}
for spin-triplet helical $p$-wave, and
\begin{align}
\hat{u}_{\Delta}&=
\begin{pmatrix}
0 & i\Delta\sin{k_y} \\
i\Delta\sin{k_y} & 0%
\end{pmatrix}%
, \\
\hat{t}_{\Delta}&=\bar{t}_{\Delta}=
\begin{pmatrix}
0 & i\frac{\Delta}{2} \\
i\frac{\Delta}{2} & 0%
\end{pmatrix}%
,
\end{align}
for chiral $p$-wave pairings.  In the normal metal $\hat{H}_\mathrm{NM}$,
the local $\tilde{u}_\mathrm{NM}(k_y)$ and nonlocal terms $\tilde{t}_\mathrm{%
NM}(k_y)$ are:
\begin{align}
\tilde{u}_\mathrm{NM}(k_y)&=
\begin{pmatrix}
\hat{u}_{0}(k_y) & 0 \\
0 & -\hat{u}^{*}_{0}(-k_y)%
\end{pmatrix}%
, \\
\tilde{t}_\mathrm{NM}(k_y)&=
\begin{pmatrix}
\hat{t}_{0}(k_y) & 0 \\
0 & -\hat{t}^{*}_{0}(-k_y)%
\end{pmatrix}%
.
\end{align}
The tunneling Hamiltonian is defined as
\begin{align}
\hat{H}_\mathrm{J}&=\hat{t}_{J}\otimes\hat{\tau}_{3},
\end{align}
with the Pauli matrices in Nambu space $\hat{\tau}_{i=0,1,2,3}$, and
\begin{align}
\hat{t}_\mathrm{J}&=-t_{1}\hat{\sigma}_{0},
\end{align}
along the $x$-direction.

We calculate the tunneling conductance based on the Lee-Fisher formula~\cite%
{LeeFisher,OhashiPRB,Fukaya_arXiv06}:
\begin{align}
\sigma^{e}_\mathrm{SC}(eV)&=\frac{1}{2\pi}\int^{\pi}_{-\pi}dk_y\bar%
\sigma^{e}(k_y,eV),
\end{align}
with
\begin{align}
\bar\sigma^{e}(k_y,eV)=\mathrm{Tr^{\prime }}[\tilde{g}_{1}(k_y,eV)+\tilde{g}%
_{2}(k_y,eV)-\tilde{g}_{3}(k_y,eV)-\tilde{g}_{4}(k_y,eV)].
\end{align}
For the spin-resolved cases, we calculate
\begin{align}
\sigma^{e}_{\uparrow}(eV)&=\frac{1}{2\pi}\int^{\pi}_{-\pi}dk_y\bar%
\sigma^{e}_\uparrow(k_y,eV), \\
\sigma^{e}_{\downarrow}(eV)&=\frac{1}{2\pi}\int^{\pi}_{-\pi}dk_y\bar%
\sigma^{e}_\downarrow(k_y,eV),
\end{align}
with
\begin{align}
\bar\sigma^{e}_\uparrow(k_y,eV)&=\tilde{g}^{\uparrow\uparrow}_{1}(k_y,eV)+%
\tilde{g}^{\uparrow\uparrow}_{2}(k_y,eV)-\tilde{g}^{\uparrow%
\uparrow}_{3}(k_y,eV)-\tilde{g}^{\uparrow\uparrow}_{4}(k_y,eV), \\
\bar\sigma^{e}_\downarrow(k_y,eV)&=\tilde{g}^{\downarrow%
\downarrow}_{1}(k_y,eV)+\tilde{g}^{\downarrow\downarrow}_{2}(k_y,eV)-\tilde{g%
}^{\downarrow\downarrow}_{3}(k_y,eV)-\tilde{g}^{\downarrow%
\downarrow}_{4}(k_y,eV).
\end{align}
For $E=eV$ with the energy $E$ and the bias voltage $V$, $\tilde{g}_{1}(k,eV)
$, $\tilde{g}_{2}(k,eV)$, $\tilde{g}_{3}(k,eV)$, and $\tilde{g}_{4}(k,eV)$
are calculated by
\begin{align}
\tilde{g}_{1}(k_y,E)&=\tilde{G}_{1,1}(k_y,E)\bar{t}^{\dagger}_\mathrm{J}%
\tilde{G}_{0,0}(k_y,E)\bar{t}_{J}, \\
\tilde{g}_{2}(k_y,E)&=\tilde{G}_{0,0}(k_y,E)\bar{t}_\mathrm{J}\tilde{G}%
_{1,1}(k_y,E)\bar{t}^{\dagger}_{J}, \\
\tilde{g}_{3}(k_y,E)&=\tilde{G}_{0,1}(k_y,E)\bar{t}^{\dagger}_\mathrm{J}%
\tilde{G}_{0,1}(k_y,E)\bar{t}^{\dagger}_{J}, \\
\tilde{g}_{4}(k_y,E)&=\tilde{G}_{1,0}(k_y,E)\bar{t}_\mathrm{J}\tilde{G}%
_{1,0}(k_y,E)\bar{t}_{J}.
\end{align}
Then $\bar{t}_\mathrm{J}$ is defined by
\begin{align}
\bar{t}_\mathrm{J}=-\hat{t}_\mathrm{J}\otimes\hat{\tau}_{0},
\end{align}
to calculate the tunneling conductance.  $\tilde{G}_{0,0}(k,E)$, $\tilde{G}%
_{1,1}(k,E)$, $\tilde{G}_{0,1}(k,E)$, and $\tilde{G}_{1,0}(k,E)$ are
obtained by
\begin{align}
\tilde{G}_{0,0}(k_y,E)&=-\frac{i}{2}[\tilde{G}^\mathrm{A}_{0,0}(k_y,E)-%
\tilde{G}^\mathrm{R}_{0,0}(k_y,E)], \\
\tilde{G}_{1,1}(k_y,E)&=-\frac{i}{2}[\tilde{G}^\mathrm{A}_{1,1}(k_y,E)-%
\tilde{G}^\mathrm{R}_{1,1}(k_y,E)], \\
\tilde{G}_{0,1}(k_y,E)&=-\frac{i}{2}[\tilde{G}^\mathrm{A}_{0,1}(k_y,E)-%
\tilde{G}^\mathrm{R}_{0,1}(k_y,E)], \\
\tilde{G}_{1,0}(k_y,E)&=-\frac{i}{2}[\tilde{G}^\mathrm{A}_{1,0}(k_y,E)-%
\tilde{G}^\mathrm{R}_{1,0}(k_y,E)].
\end{align}
Here, $\hat{G}^\mathrm{R}_{i,j}(k_y,E)$ and $\hat{G}^\mathrm{A}_{i,j}(k_y,E)$
denote the retarded (R) and advanced (A) Green's function with the indices $%
i,j=0,1$.  Local $\tilde{G}_{0,0}(k_y,E)$ and $\tilde{G}_{1,1}(k_y,E)$, and
nonlocal Green's functions $\tilde{G}_{0,1}(k_y,E)$ and $\tilde{G}%
_{1,0}(k_y,E)$ are calculated by
\begin{align}
\tilde{G}_{00}(k_y,z)&=[\{\tilde{G}^\mathrm{L}_{0}(k_y,z)\}^{-1}-\tilde{t}_%
\mathrm{J}\tilde{G}^\mathrm{R}_{1}(k_y,z)\tilde{t}^{\dagger}_\mathrm{J}%
]^{-1}, \\
\tilde{G}_{11}(k_y,z)&=[\{\tilde{G}^\mathrm{R}_{1}(k_y,z)\}^{-1}-\tilde{t}%
^{\dagger}_\mathrm{J}\tilde{G}^\mathrm{L}_{0}(k_y,z)\tilde{t}_\mathrm{J}%
]^{-1}, \\
\tilde{G}_{01}(k_y,z)&=\tilde{G}^\mathrm{L}_{0}(k_y,z)\tilde{t}_\mathrm{J}%
\tilde{G}_{11}(k_y,z), \\
\tilde{G}_{10}(k_y,z)&=\tilde{G}^\mathrm{R}_{1}(k_y,z)\tilde{t}^{\dagger}_%
\mathrm{J}\tilde{G}_{00}(k_y,z),
\end{align}
where $z=E\pm i\delta$ with the infinitesimal value $\delta=0.01\Delta$ are
the retarded (R) $z=E+i\delta$ and advanced (A) parts $z=E-i\delta$,
respectively.  For the recursive Green's function method~\cite{Umerski}, we
obtain the surface Green's functions:
\begin{align}
\tilde{G}^\mathrm{L}_{0}(k_y,z)&=[z-\tilde{u}_\mathrm{SC}(k_y)-\tilde{t}%
^{\dagger}_\mathrm{SC}\tilde{G}^\mathrm{L}(k_y,z)\tilde{t}_\mathrm{SC}]^{-1},
\\
\tilde{G}^\mathrm{R}_{1}(k_y,z)&=[z-\tilde{u}_\mathrm{NM}(k_y)-\tilde{u}_%
\mathrm{B}-\tilde{t}_\mathrm{NM}\tilde{G}^\mathrm{R}(k_y,z)\tilde{t}%
^{\dagger}_\mathrm{NM}]^{-1},
\end{align}
with the semi-infinite Green's functions $\tilde{G}^\mathrm{L}(k_y,z)$ and $%
\tilde{G}^\mathrm{R}(k_y,z)$, and $\tilde{u}_\mathrm{B}=U_{b}\hat{\sigma}%
_{0}\otimes\hat{\tau}_{3}$ for the strength of the potential barrier $U_b=5t$%
.  We first calculate the semi-infinite Green's functions~\cite{Umerski} for
both retarded ($z=E+i\delta$) and advanced ($z=E-i\delta$) parts to obtain
the tunneling conductance.  \newline

\noindent \textbf{S2. Additional results in the tight-binding model}\newline

From now on, we show the results for the tunneling conductance in the
tight-binding model.  Here, we focus on spin-singlet $d_{x^2-y^2}$, $d_{xy}$%
-wave, chiral $d$-wave and spin-triplet chiral $p$-wave pairings as well as $%
d_{xy}$ and $d_{x^2-y^2}$-wave altermagnetism.  In Fig.~\ref%
{fig:dx2_y2_wave_dxy_AM}, we plot the momentum-resolved tunneling
conductance [Figs.~\ref{fig:dx2_y2_wave_dxy_AM} (a)-(d)] and the tunneling
conductance [Figs.~\ref{fig:dx2_y2_wave_dxy_AM} (e)-(h)] in $d_{x^2-y^2}$%
-wave superconductors with $d_{xy}$-wave altermagnetism/normal metal
junctions at (a)(e) $t_{xy}=0$, (b)(f) $t_{xy}=0.5\Delta$, (c)(g) $%
t_{xy}=\Delta$, (d)(h) $t_{xy}=1.5\Delta$.  Here, $\sigma_\mathrm{N}$
denotes the tunneling conductance at zero energy in the normal state.
Without any altermagnetism at $t_{xy}=0$, nodal structure of $d_{x^2-y^2}$%
-wave superconductors appears [Fig.~\ref{fig:dx2_y2_wave_dxy_AM} (a)] and
the resulting conductance becomes V-shaped [Fig.~\ref{fig:dx2_y2_wave_dxy_AM}
(e)].  In the presence of the $d_{xy}$-wave altermagnetism, the energy gap
closes [Figs.~\ref{fig:dx2_y2_wave_dxy_AM} (b)(c)(d)], and the value of the
tunneling conductance at zero energy becomes larger with the increase of $%
t_{xy}$ [Figs.~\ref{fig:dx2_y2_wave_dxy_AM} (f)(g)(h)].   When we focus on the $%
d_{x^2-y^2}$-wave altermagnetism $t_{x^2-y^2}$ shown in Figs.~\ref%
{fig:dx2_y2_wave_dx2_y2_AM} (a)(d)(e)(f)(g)(h)(i)(l), we also obtain the
same results as those in the continuum model shown in Fig.\ (b) of the
main text.  Then the spin-resolved tunneling conductance $%
\sigma^{\uparrow,\downarrow}_\mathrm{SC}(eV)$ satisfies $\sigma^{\downarrow}_%
\mathrm{SC}(eV)=\sigma^{\uparrow}_\mathrm{SC}(-eV)$ [Figs.~\ref%
{fig:dx2_y2_wave_dx2_y2_AM} (d)(h)(l)] as well as the momentum and
spin-resolved tunneling conductance: $\bar{\sigma}_{\downarrow}(k_y,eV)=\bar{%
\sigma}_{\uparrow}(k_y,-eV)$ [Figs.~\ref{fig:dx2_y2_wave_dx2_y2_AM}
(b)(c)(f)(g)(j)(k)].

In the case of $d_{xy}$-wave pairings, the momentum-resolved tunneling
conductance shown in Figs.~\ref{fig:dxy_wave_dxy_AM} (a)-(d) remains the
zero-energy flat bands of the surface Andreev bound states for $t_{xy}<\Delta
$, and they disappear for $t_{xy}\ge\Delta$.  These behaviors correspond to
Fig.\ 2 (d) of the main text.  We obtain the same line shapes of the
tunneling conductance [Figs.~\ref{fig:dxy_wave_dxy_AM} (e)-(h)] as those in
the continuum model [Fig.\ 2 (d) in the main text].  With the $d_{x^2-y^2}$%
-wave altermangetism $t_{x^2-y^2}$, the zero energy flat bands of the
surface Andreev bound states split and they cross at finite momentum [Figs.~%
\ref{fig:dxy_wave_dx2_y2_AM} (a)(e)(i)] and this behavior coincides with
Fig.\ 2 (c) in the main text.  When we see the momentum and spin-resolved
tunneling conductance $\bar{\sigma}_{\uparrow,\downarrow}(k_y,eV)$ in Figs.~%
\ref{fig:dxy_wave_dx2_y2_AM} (b)(c)(f)(g)(j)(k), the spin-split surface
Andreev bound states appear.  As a result, the zero energy peak of the
tunneling conductance collapses shown in Fig.\ 2 (c) of the main text, and
then the peak of the spin-resolved tunneling conductance shifts with the
increase of the $d_{x^2-y^2}$-wave altermagnetism [Figs.~\ref%
{fig:dxy_wave_dx2_y2_AM} (d)(h)(l)].  As well as these $d$-wave pairings, we obtain the same behavior of the tunneling
conductance in the chiral $p$ and $d$-wave superconductors with $d$-wave
altermagnetism [Fig.~\ref{fig:chiral_p_wave_dxy_AM}, Fig.~\ref%
{fig:chiral_p_wave_dx2_y2_AM}, Fig.~\ref{fig:chiral_d_wave_dxy_AM}, and Fig.~%
\ref{fig:chiral_d_wave_dx2_y2_AM}]. They are consistent with Figs.\ 3 and 4 in the main text.  While, in the helical $p$-wave pairing
shown in Fig.~\ref{fig:helical_p_wave_d_AM}, the resulting tunneling
conductance does not change in the presence of the large $d$-wave
altermagnetism at $t_{xy},t_{x^2-y^2}=10\Delta$ and $40\Delta$.  Because the
Cooper pair can form at $k$ and $-k$ with the same spin in helical $p$-wave
superconductors with the $d$-wave altermagnetism, the tunneling conductance
is robust with the altermagnetism.  \newline

\begin{figure}[t]
\begin{center}
\includegraphics[width=0.5\linewidth]{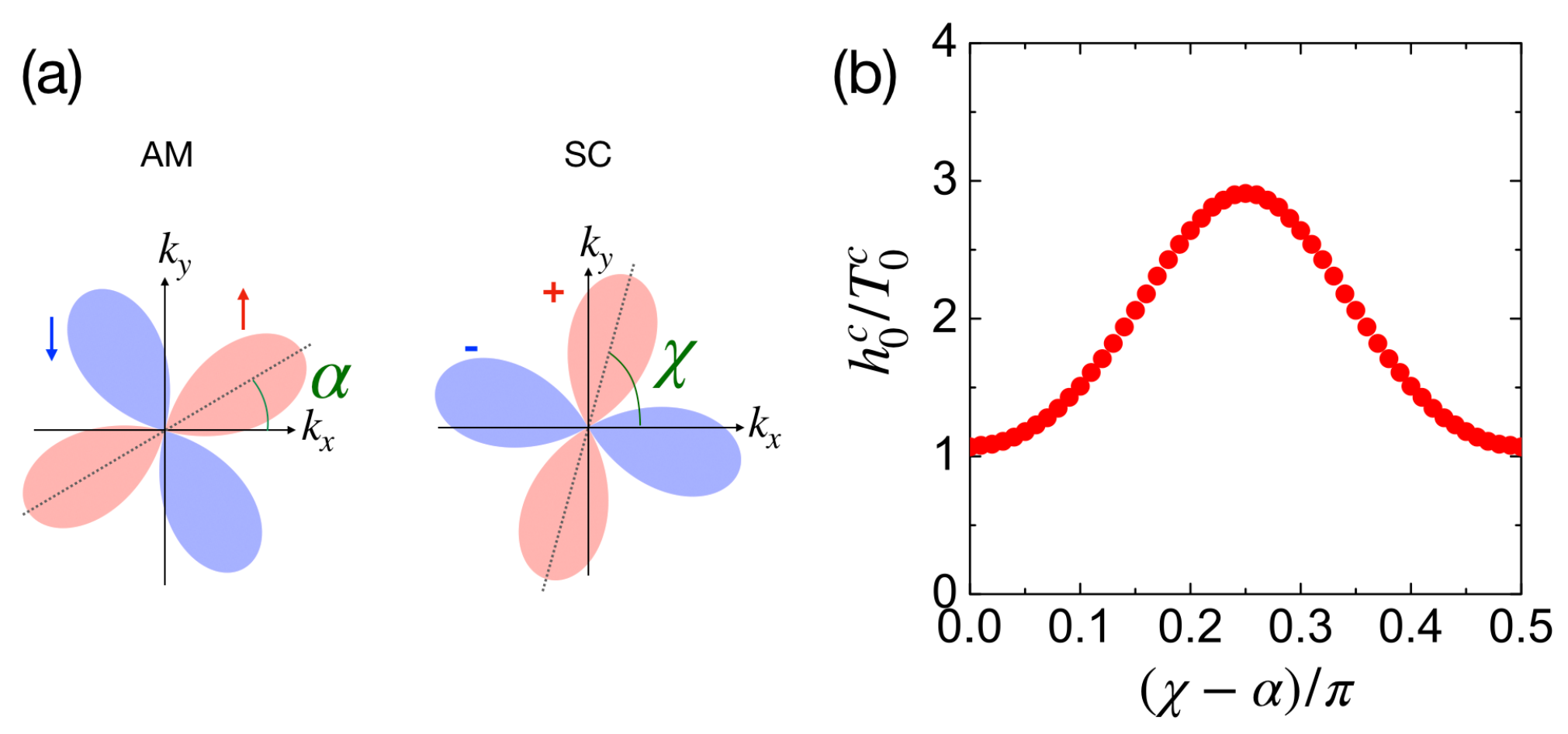}
\end{center}
\par
\caption{(a) Schematics of $d$-wave AM and $d$-wave SC order
parameters, where $\protect\alpha$ and $\protect\chi$ denotes the AM and SC
orientations, respectively. (b) Critical altermagnetic field $h_{0}^{c}$
dependence on the twist angle $\protect\chi-\protect\alpha$ for an AM-SC
bilayer with a $d$-wave SC.}
\label{criticalh}
\end{figure}

\noindent \textbf{S3. Details of the self-consistent
calculation of the critical field}\newline 

Here, we address the possibility of the coexistence of $d$-wave AM and $d$%
-wave SC following the pioneering work in altermagnetic superconductor \cite%
{Belzig}. The superconducting order parameter is calculated
self-consistently using Eilenberger equation
\begin{equation}
\left[ \omega _{n}\hat{\sigma}_{0}\hat{\tau}_{z}+\check{\Delta}(\hat{k})-ih(
\hat{k})\hat{\sigma}_{z}\hat{\tau}_{z},\check{g}_{\omega _{n}}(\hat{k}) %
\right] =0,
\end{equation}
where $\omega _{n}=(2n+1)\pi T\,$is the Matsubara frequency with $n$ an
integer. For $d$-wave pairings, the self-consistency equation is given by
\begin{eqnarray}
1 &=&\pi \lambda T\sum\limits_{0<\omega _{n}<\omega _{c}}\int_{0}^{2\pi }
\frac{d\theta }{\pi }\frac{\cos ^{2}2\left( \theta -\chi \right) }{\sqrt{
\left( \omega _{n}+ih\cos 2\theta \right) ^{2}+\Delta ^{2}\cos ^{2}2\left(
\theta -\chi \right) }}  \notag \\
&&+\pi \lambda T\sum\limits_{0<\omega _{n}<\omega _{c}}\int_{0}^{2\pi }\frac{
d\theta }{\pi }\frac{\cos ^{2}2\left( \theta -\chi \right) }{\sqrt{\left(
\omega _{n}-ih\cos 2\theta \right) ^{2}+\Delta ^{2}\cos ^{2}2\left( \theta
-\chi \right) }}.
\end{eqnarray}
Here, we have set $\alpha=0$ so that $\chi $ is the twisted angle between SC
and AM. $\omega _{c}$ is  the cut-off frequency. To determine the critical
field as a function of  temperature $T$, we assume $\Delta \rightarrow 0$
and calculate the critical  field $h_{c}(T)$ satisfying
\begin{eqnarray}
1 &=&\pi \lambda T\sum\limits_{0<\omega _{n}<\omega _{c}}\int_{0}^{2\pi }
\frac{d\theta }{\pi }\frac{\cos ^{2}2\left( \theta -\chi \right) }{\omega
_{n}+ih_{c}\cos 2\theta } +\pi \lambda T\sum\limits_{0<\omega _{n}<\omega
_{c}}\int_{0}^{2\pi }\frac{ d\theta }{\pi }\frac{\cos ^{2}2\left( \theta
-\chi \right) }{\omega _{n}-ih_{c}\cos 2\theta }.
\end{eqnarray}
We can use $\lambda $ as a function of $\omega _{c}$ and $T$, which can
reproduce the well-know self-consistent result under zero field.

We show the critical altermagnetic strength $h_{0}^{c}$ as a  function of
the twist angle between AM and SC in Fig.\thinspace \ref{criticalh}. We set
the temperature to be sufficiently low $ T=0.02T_{0}^{c}$, with $T_{0}^{c}$
the transition temperature of a $d$ -wave SC without altermagnetism. It is
seen that SC states can survive at an arbitrary  twist angle unless the
field strength $h_{0}$ exceeds $h_{0}^{c}$. When the twist angle is $0$ ({$%
\pi /4$}), the critical field  develops a minimum (maximum) value. The
maximum critical field indicates a favorable coexistence of $d$-wave SC and $%
d$-wave AM at a twist $\pi /4$, where vanishing AM field and maximum $d$%
-wave pair potential aligns in the same orientation. For the same reason,
the minimum critical field occurs at no tiwst since the AM field and $d$%
-wave pairing grow simultaneously and compete each other. In the main text,
we consider a small altermagnetic field induced on the SC side because the
inverse proximity effect suppress the AM field significantly by the
tunneling rates across the AM-SC interface, and can satisfy $h_{0}<h_{0}^{c}$%
.

For chiral $p$- or chiral $d$-wave pairings, we regard that they have the
same phase diagram with $s$-wave pairings \cite{Belzig} since the bulk state
is fully  gapped and their momentum-dependent phase shows no role to
determine  the amplitude of order parameter. It is noted that our assumption
can be  applied to the bulk system with hybrid SC and AM. For
heterojunctions, the  order parameters will alter spatially and it is a
future task to solve  Eilenberger equation involving AM in the junction
systems.\newline

\begin{figure}[t]
\begin{center}
\includegraphics[width=0.8\linewidth]{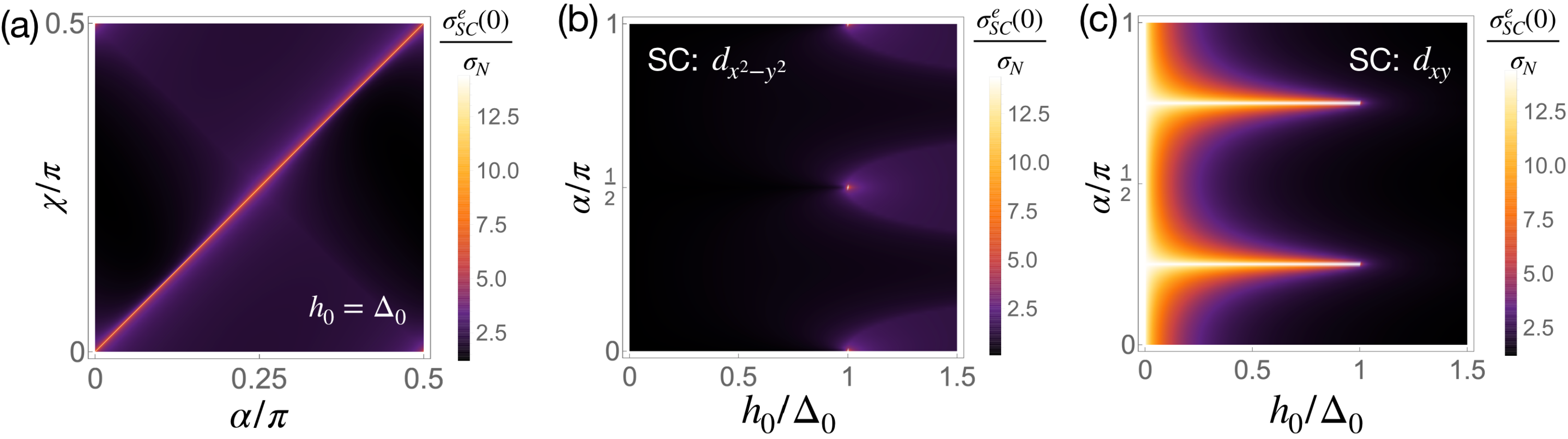}
\end{center}
\par
\caption{(a) Zero bias conductance of $d$-wave SC with $d$-wave
AM as a function of $\protect\chi$ and $\protect\alpha$ at $h_0=\Delta_0$.
(b,c) Zero bias conductance vs $\protect\alpha$ and $h_0$ for a $%
d_{x^{2}-y^{2}}$-wave SC (c) and a $d_{xy}$-wave SC (d). We set $Z=2$.} 
\label{Fig3}
\end{figure}

\begin{figure}[t]
\begin{center}
\includegraphics[width=0.6\linewidth]{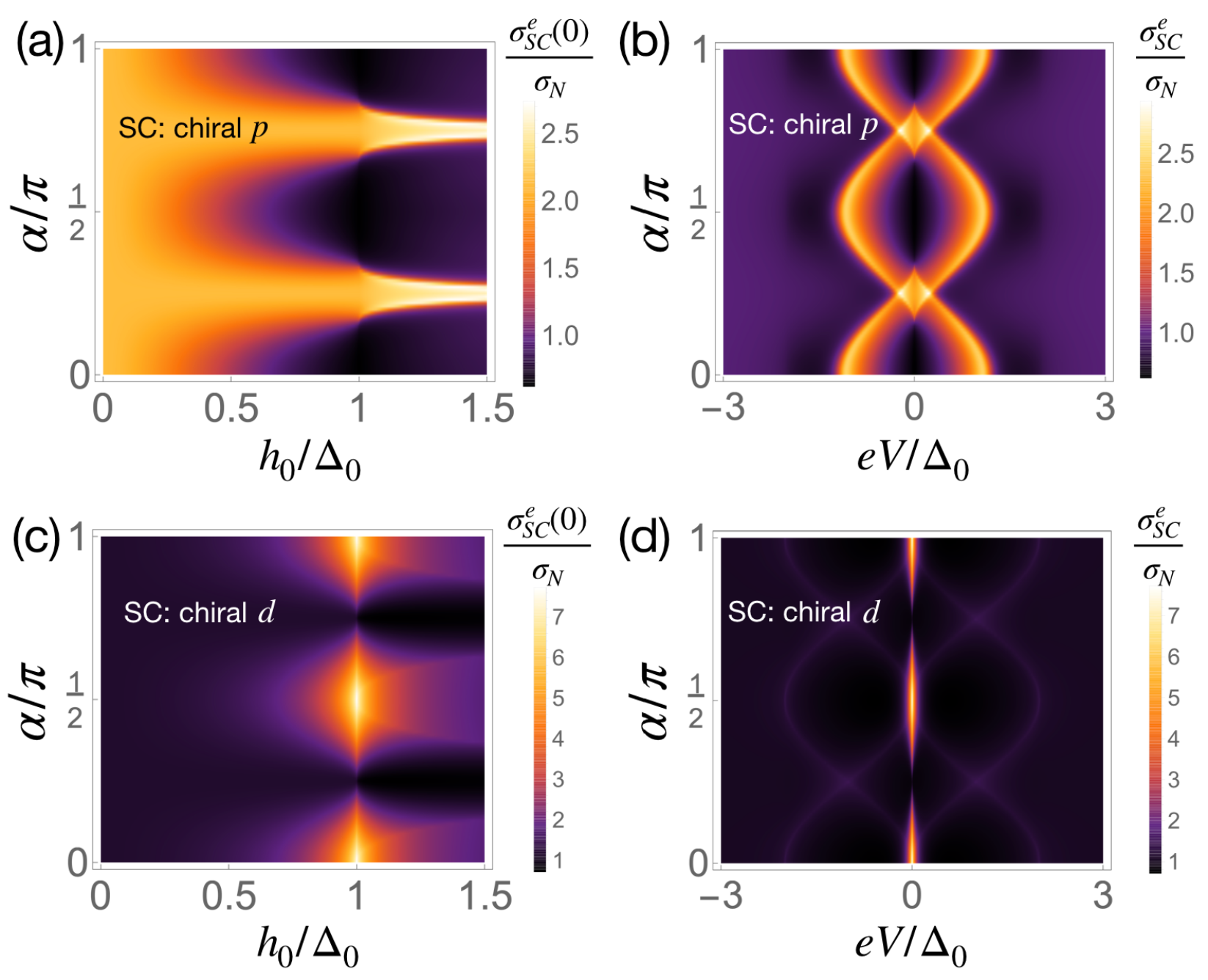}
\end{center}
\par
\caption{ (a) Zero bias conductance vs $\protect\alpha$ and $h_0
$ for a chiral $p$-wave SC with a $d$-wave AM, where $\protect\alpha$ is the
orientation of AM. (b) Same junction but the total conductance as a function
of $eV$ and $\protect\alpha$ at $h_0=\Delta_0$. (c)(d) are the same with
(a)(b), but for a chiral $d$-wave SC. We set $Z=2$ for both panels.} 
\label{twistchiral}
\end{figure}

\noindent \textbf{\textbf{S4.} Influence of crystal misalignment} \newline

We first discuss the influence of a twist angle between a $d$-wave AM and a $%
d$-wave SC on the conductance spectrum. We plot the zero-biased conductance
(ZBC) as a function of $\chi $ and $\alpha $ at $h_{0}=\Delta _{0}$ in Fig.~%
\ref{Fig3}(a). It is seen that a robust ZBCP forms at $\chi =\alpha $ as a
result of both BFS ($\chi =0$, $d_{x^{2}-y^{2}}$-wave SC) and SABSs ($\chi
\neq 0$, $d_{xy}$-wave SC). For non-zero twist $\chi \neq \alpha $, such
enhancement of conductance at zero voltage largely vanishes. For $h_{0}\neq
\Delta _{0}$, we can see from Fig.\thinspace \ref{Fig3}(b) that the $%
d_{x^{2}-y^{2}}$-wave SC produces a ZBCP at $h_{0}=\Delta _{0}$ and with a  $%
d_{x^{2}-y^{2}}$-wave AM ($\alpha =0$ or $\pi /2$). However, the ZBCP is
robust across the region $0\leq {h_{0}}\leq \Delta _{0}$ for a $d_{xy}$%
-wave SC with a $d_{xy}$-wave AM ($\alpha =\pi /4$ or $3\pi /4$) [Fig.\thinspace \ref{Fig3}(c)].

In the main text, we have shown the conductance for a chiral $p$-wave SC
with a $d_{x^{2}-y^{2}}$-wave AM in Fig.3(c) and a chiral $p$-wave SC with a
$d_{xy}$-wave AM in Fig.3(d). The ZBC is greatly suppressed in the $%
d_{x^{2}-y^{2}}$-wave AM case due to the spin splitting of SABSs. In Fig.\thinspace \ref {twistchiral}(a), we show the evolution of ZBC with arbitrary orientation of AM. It is
seen that for $\alpha =\pi /4$, $3\pi /4$ ($d_{xy}$-wave AM), ZBC reaches
its maximum value while it decreases for other values of $\alpha $. The
double-peak structure for a large value of $h_{0}$ can also be obtained for
any angle $\alpha $, as seen in Fig.\thinspace \ref {twistchiral}(b) where we set $h_{0}=\Delta _{0}$.
We now turn to discuss the chiral $d$-wave SC with a twisted AM. It has been
shown in Fig.4(c) in the main text that with a $d_{x^{2}-y^{2}}$-wave AM, a
ZBCP develops due to the flattened chiral bands at $h_{0}=\Delta _{0}$. In
Fig.\thinspace \ref {twistchiral}(c), we show the ZBC as a function of AM orientation $\alpha $ and the
altermagnetic field $h_{0}$. It is found that the ZBCP\ appears for large
parameter regions where $\alpha $ is closed to $0$ or $\pi /2$ and $h_{0}$
is closed $\Delta _{0}$.  Figure \thinspace \ref {twistchiral} (d) further show that the ZBC\ forms a peak
at zero voltage for a large parameter region of $\alpha $ when $h_{0}=\Delta
_{0}$. Such behavior demonstrates the robustness of the ZBCP found in chiral
$d$-wave SC with AM even with a twist angle. It can be understood that
except the $d_{xy}$-wave AM, the SABSs will split and one of the SABSs will
be more flattened to enhance the zero energy states, resulting in the
enhanced ZBC.\newline

\noindent \textbf{\textbf{S5.} Subgap states for superconductors with $g$-wave altermagnetism} \newline

\begin{figure}[tbh]
    \centering
    \includegraphics[width=0.8\linewidth]{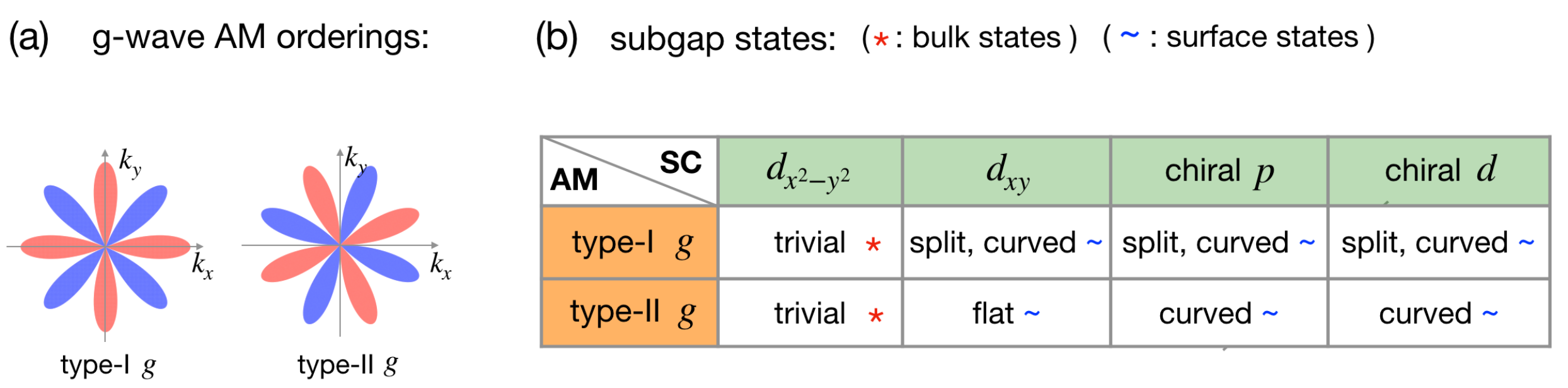}
\caption{(a) Illustration of $g$-wave altermagnet orderings. (b) Induced subgap states in superconductors with $g$-wave altermagnet. }
\label{gwavedemo}
\end{figure}

In this section, we consider AMs exhibiting $g$-wave
parity {\color{blue}[Fig.\,\ref{gwavedemo}(a)]}, 
expected, e.g., in MnTe.   
We focus on the $g$-wave altermgnetic term $m_{\bm{k}}^{g1}=(\hat{k}_{x}^{2}-\hat{k}_{y}^{2})
^{2}-4\hat{k}_{x}^{2}\hat{k}_{y}^{2}$ and $m_{\bm{k}}^{g2}=4%
\hat{k}_{x}\hat{k}_{y}(\hat{k}_{x}^{2}-\hat{k}_{y}^{2})$, denoted as type-I and type-II $g$-wave AMs here. In Fig.\,\ref{gwavedemo}(b), we show the types of subgap states induced by $g$-wave AM and discuss separately in the followings. It is noted that all the methods of analysis in the main text are still valid here for the $g$-wave AM. 

\begin{figure}[tbh]
    \centering
    \includegraphics[width=0.6\linewidth]{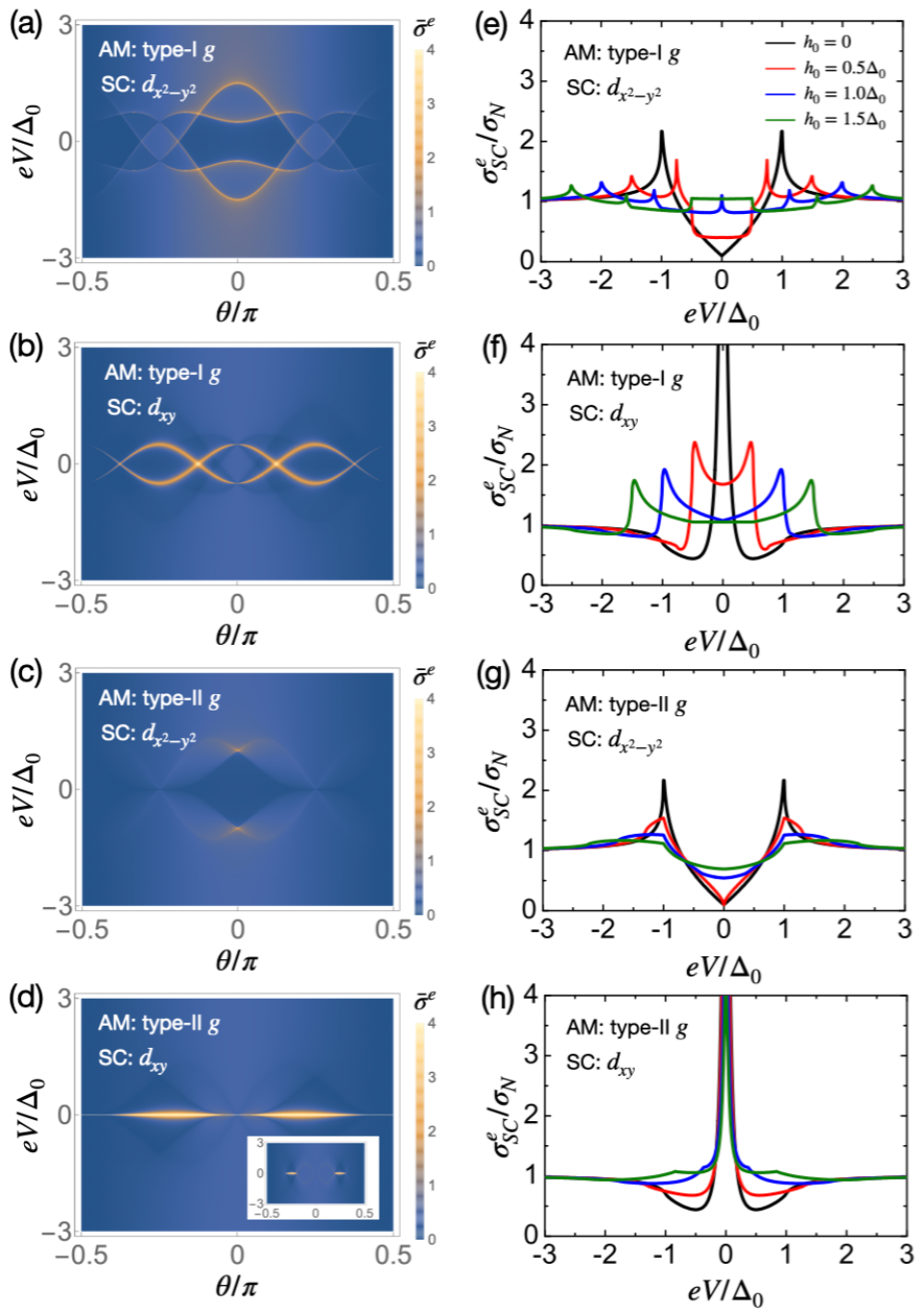}
\caption{(a) Angle-resolved charge conductance for a junction having  a chiral $d_{x^{2}-y^{2}}$-wave SC  with a type-I $g$-wave AM at $h_{0}=0.5\Delta_{0}$, $Z=2$. (e) Total charge conductance for the system of (a) for distinct values of $h_{0}$.  (b,c,d) and (f,g,h) The same as in (a) and (e), respectively, but for distinct combinations  of $g$-wave AMs with $d_{x^{2}-y^{2}}$- and $d_{xy}$-wave SCs. The inset in (d) shows the same as in (d) but at $h_{0}=1.5\Delta_{0}$. }
\label{gAM_dSC}
\end{figure}

\begin{figure}[tbh]
    \centering
    \includegraphics[width=0.6\linewidth]{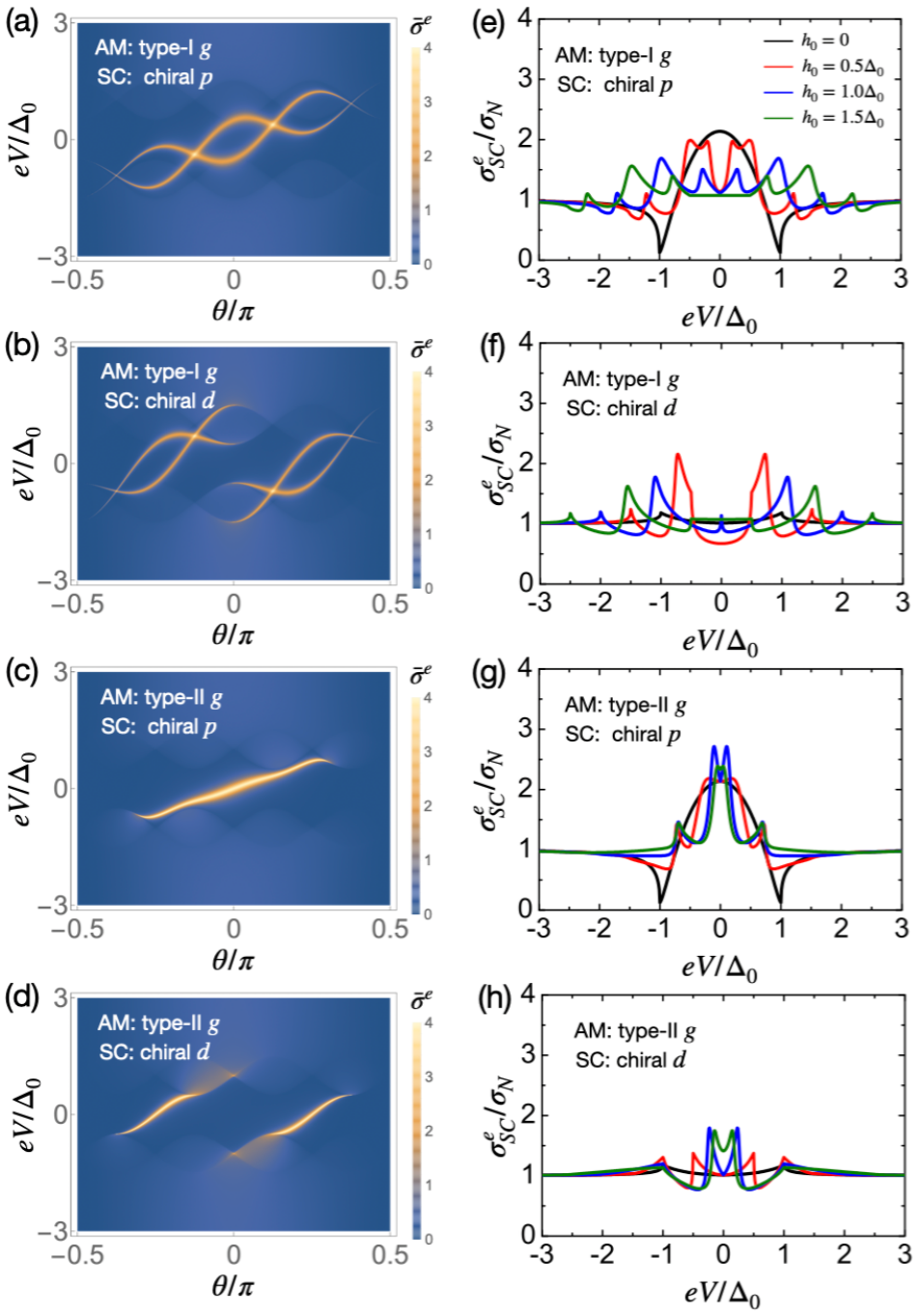}
\caption{(a) Angle-resolved charge conductance for a junction having  a chiral $p$-wave SC   with a type-I $g$-wave AM at $h_{0}=0.5\Delta_{0}$, $Z=2$. (b) Total charge conductance for the system of (a) for distinct values of $h_{0}$.  (b,c,d) and (g,h,i) The same as in (a) and (b), respectively, but for distinct combinations  of $g$-wave AMs with chiral $p$-wave and $d$-wave SCs.}
\label{gAM_chiral}
\end{figure}

First, we address the emergence of subgap states and their signatures in  hybrid systems composed of a $g$-wave AMs and a $d$-wave SC. The emergence of SABSs can be anticipated by using simple arguments. For type-I $g$-wave AMs, the anisotropic field is $m^{g1}_{\bm{k}}=(\hat{k}_{x}^{2}-\hat{k}_{y}^{2})^{2}-4\hat{k}_{x}^{2}\hat{k}_{y}^{2}$, and has even parity $m_{( k_{x},k_{y}) }=m_{( -k_{x},k_{y}) }$, as for $d_{x^{2}-y^{2}}$-wave AMs. Thus, we can expect spin-split SABSs since  $\Gamma _{\uparrow }^{+}\Gamma _{\uparrow }^{-}\neq \Gamma_{\downarrow }^{+}\Gamma _{\downarrow }^{-}$, see Eq.(7) in the main text. For type-II $g$-wave AMs, the field is  $m^{g2}_{\bm{k}}=4\hat{k}_{x}\hat{k}_{y}(\hat{k}_{x}^{2}-\hat{k}_{y}^{2}) $, and the SABSs remain spin degenerate since $m_{(k_{x},k_{y})}$ changes sign with $k_{x}$ as $m_{(k_{x},k_{y}) }=-m_{(k_{x},k_{y})}$ and the relation $\Gamma _{\uparrow }^{+}\Gamma _{\uparrow }^{-}=\Gamma_{\downarrow }^{+}\Gamma _{\downarrow }^{-}$ is respected.

To demonstrate the emergence of SABSs in $g$-wave AM with $d$-wave superconductivity, in Fig.\,\ref{gAM_dSC} we show the charge conductance at $Z\neq0$.
For type-I  $g$-wave AMs, the angle-resolved charge conductance in Figs.\,\ref{gAM_dSC}(a,b) shows the emergence of curved subgap states at $h_{0}=0.5\Delta_{0}$, which are spin split and   consistent with the reasoning of the previous paragraph. When the SC has $d_{x^{2}-y^{2}}$-wave symmetry, the total charge conductance has a V-shape profile as a function of $eV$ at $h_{0}=0$, which is then filled up as $h_{0}$ increases following a U-shape dependence [Fig.\,\ref{gAM_dSC}(e)]. At  $h_{0}=\Delta_{0}$, a tiny ZBCP appears over a filled subgap conductance, but it transitions into a constant subgap conductance for larger  values of $h_{0}$  [Fig.\,\ref{gAM_dSC}(e)]. It is worth noting that for finite type-I $g$-wave altermagnetism, there appear more spikes than $d$-wave AM due to the richer nodal structures of $g$-wave AM field. In the case of a SC with $d_{xy}$-wave symmetry, the total charge conductance at zero field has a zero bias peak due to the degenerate SABSs; increasing the altermagnetic field suppresses the ZBCP into a soft U-shape conductance whose edges increase as $h_{0}$ increases [Fig.\ref{gAM_dSC}(f)]. The conductance edges are given by $eV=h_0$ and their positions can be used to determine the $g$-wave altermagnetic strength.

For a type-II $g$-wave AMs, the angle-resolved charge conductance shows that 
no split SABSs appear  at $h_{0}\neq0$, again consistent with the discussion made at the beginning of this section, see  Figs.\,\ref{gAM_dSC}(c,d). The zero-energy flat band of the $d_{xy}$-wave SC remains robust at $h_{0}\neq0$, see Fig.\,\ref{gAM_dSC}(d) and its inset. The effect of type-II $g$-wave altermagnetism in $d_{x^{2}-y^{2}}$-wave SCs is to fill up the V-shape total charge subgap conductance into a U-shape profile that is nonzero at zero bias [Fig.\,\ref{gAM_dSC}(g)]. Moreover, type-II $g$-wave altermagnetism in $d_{xy}$-wave SCs does not modify the ZBCP due to the surface zero-energy flat band, see Figs.\,\ref{gAM_dSC}(d,h). Only the conductance around the ZBCP gets filled by increasing the strength of type-II $g$-wave altermagnetism. The impact of this type of AMs on the ZBCP of $d_{xy}$-wave SCs is different from what occurs with $d$-wave AMs.

Next, we discuss the SABSs and their signatures in chiral SCs   with $g$-wave altermagnetism. 
For the SABSs in chiral SCs, our previous analysis of spin splittings is
still valid. To see this, in Figs.\,\ref{gAM_chiral}(a,b) we show the angle-resolved conductance of chiral $p$- and $d$-wave SCs with type-I $g$-wave altermagnetism.
The chiral SABSs emerge following a curved dispersion, split for segments of $\theta$ but degenerate at special angles given by $\theta=\pm \pi/8,\ \pm3\pi/8$ [Figs.\,\ref{gAM_chiral}(a,b)]. While chiral $p$-wave SCs host one pair of curved SABSs with four points of degeneracy, chiral $d$-wave SCs host two pairs of SABSs each pair with a degeneracy at the same values of $\theta$. Those crossing points where SABSs are degenerated reveal the $g$-wave nature of altermagnetism where $\mathcal{M}^{g}_{\bm{k}}=0$. 
As a result, the dispersion of the chiral surface states reaches  zero energy multiple times and develops quasi-flat surface states inside the gap.  This property of surface states gives rise  to the spikes of the total charge conductance of Figs.\ref{gAM_chiral}(e,f).

For type-II $g$-wave AMs,  the chiral states  are spin degenerate and acquire a soft curved dispersion, which is shown by the angle-resolved charge conductance in  Figs.\ref{gAM_chiral}(c,d). These curved chiral states also develop spikes in the total charge conductance [Figs.\ref{gAM_chiral}(g,h)],  where the  chiral $p$-wave SCs with type-II $g$-wave AMs are similar to $d_{xy}$-AM, while the chiral $d$-wave SCs with type-II $g$-wave AMs has no surface flat bands unlike $d_{x^{2}-y^{2}}$-wave AMs.\newline

\noindent \textbf{\textbf{S6.} Tailoring spin conductance by
altermagnetism in unconventional SCs} \newline

\begin{figure}[t]
\centering
\includegraphics[width=0.5\linewidth]{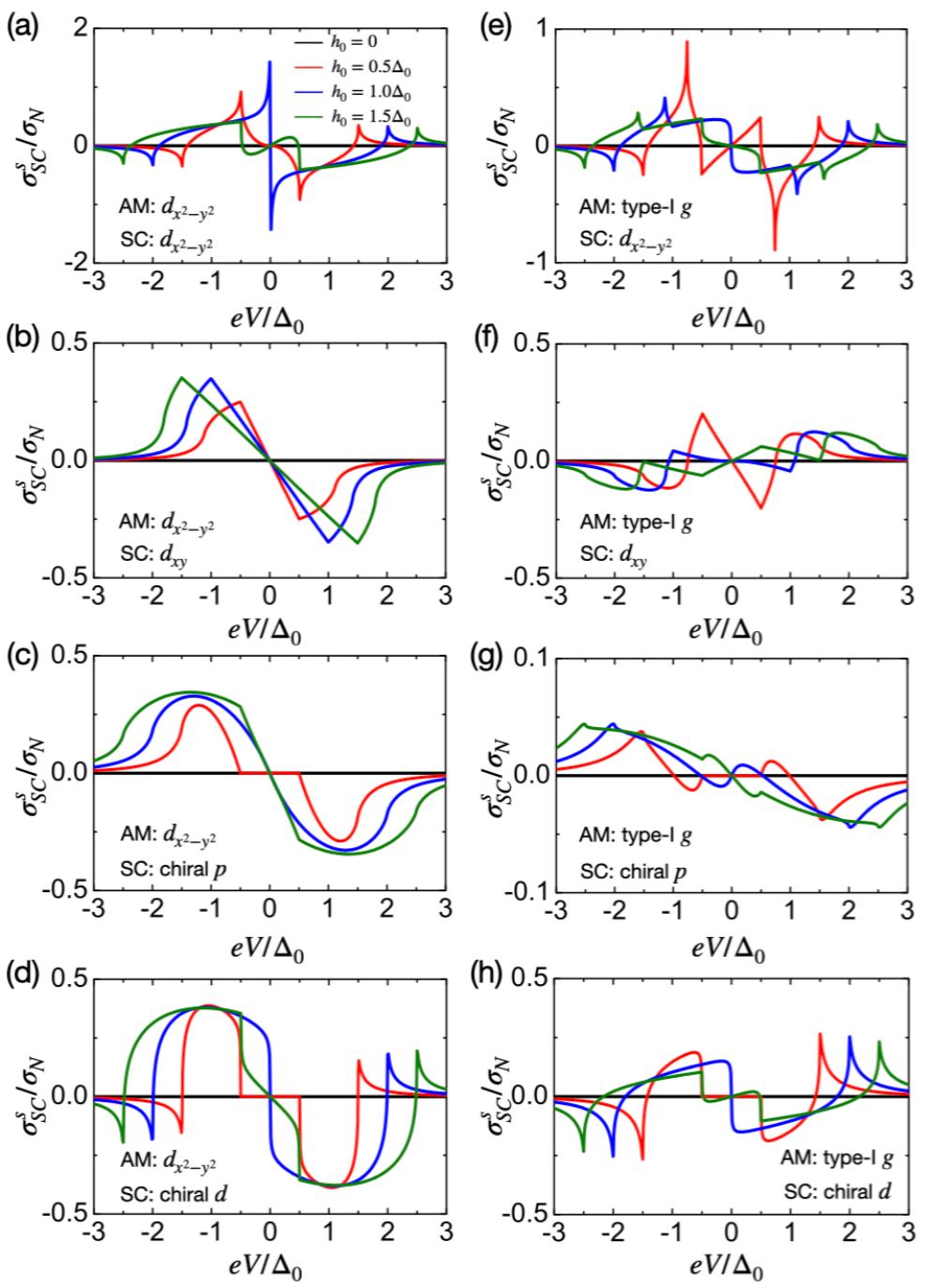}
\caption{Spin conductance as a function of $eV$ for various pair potentials
and types of AMs, including $d_{x^2-y^2}$-wave AMs (a-d) and type-I $g$-wave
AMs (e-h). We set the barrier strength $Z=3$ for all panels.}
\label{spincond}
\end{figure}

Here, we explore the application of our AM-SC system by  focusing on the
spin conductance $\sigma _{\mathrm{SC}}^{s}$ which measures  the ability to
produce a spin current and plays a key role in  superconducting spintronics.
The angle-resolved spin conductance is found as  \cite{spincur1,spincur2}
$\bar{\sigma}_{\sigma }^{s}=1-|a_{\sigma }|^{2}-|b_{\sigma }|^{2}$, which
gives
\begin{equation}
\bar{\sigma}_{\sigma }^{s}=\bar{\sigma}_{n}\frac{1-\bar{\sigma}_{n}|\Gamma
_{\sigma }^{+}|^{2}+(\bar{\sigma}_{n}-1)|\Gamma _{\sigma }^{+}\Gamma
_{\sigma }^{-}|^{2}}{|\Lambda -\eta (1-\bar{\sigma}_{n})\Gamma _{\sigma
}^{+}\Gamma _{\sigma }^{-}|^{2}}\,,  \label{ARSC}
\end{equation}
The denotations are the same with those in the main text. Then, the spin
conductance is calculated as
\begin{equation}
\sigma _{\mathrm{SC}}^{s}=\frac{e^{2}}{h}\int_{-\frac{\pi }{2}}^{\frac{\pi }{
2}}\bar{\sigma}^{s}\cos \theta d\theta ,  \label{SCond}
\end{equation}
where $\bar{\sigma}^{s}=\bar{\sigma}_{\uparrow }^{s}-\bar{\sigma}
_{\downarrow }^{s}$ is the angle-resolved spin conductance.

In general, we stress that the non-zero spin conductance originates from the
spin splitting due to altermagnetism with $d_{x^{2}-y^{2}}$-wave and type-I $
g$-wave symmetries. Considering these AMs, in Fig.\thinspace \ref{spincond}
we show the spin conductance with SCs corresponding to the main text. For a $
d_{xy}$-wave SC and chiral $p$-wave SC, both with $d_{x^{2}-y^{2}}$-wave AM,
the spin conductance exhibits only one sign flip as $eV$ changes from
negative to positive values [Figs.\ref{spincond}(b,c)]. A chiral $d$-wave SC
with $d_{x^{2}-y^{2}}$-wave AM develops three sign changes, including a fast
change at $eV=0$, similar but less pronounced than for $d_{x^{2}-y^{2}}$
-wave SCs in Fig.\thinspace \ref{spincond}(a). Depending on the type of $g$
-wave altermagnetism and SCs, the spin conductance acquires multiple changes
of its sign, with features that are entirely tied to the type of AM. Thus,
strong dependence of the spin conductance on the type of altermagnetism and
its strength offers a robust way for manipulating spin currents in
superconducting platforms.\newline

\begin{figure}[t]
\begin{center}
\includegraphics[width=0.6\linewidth]{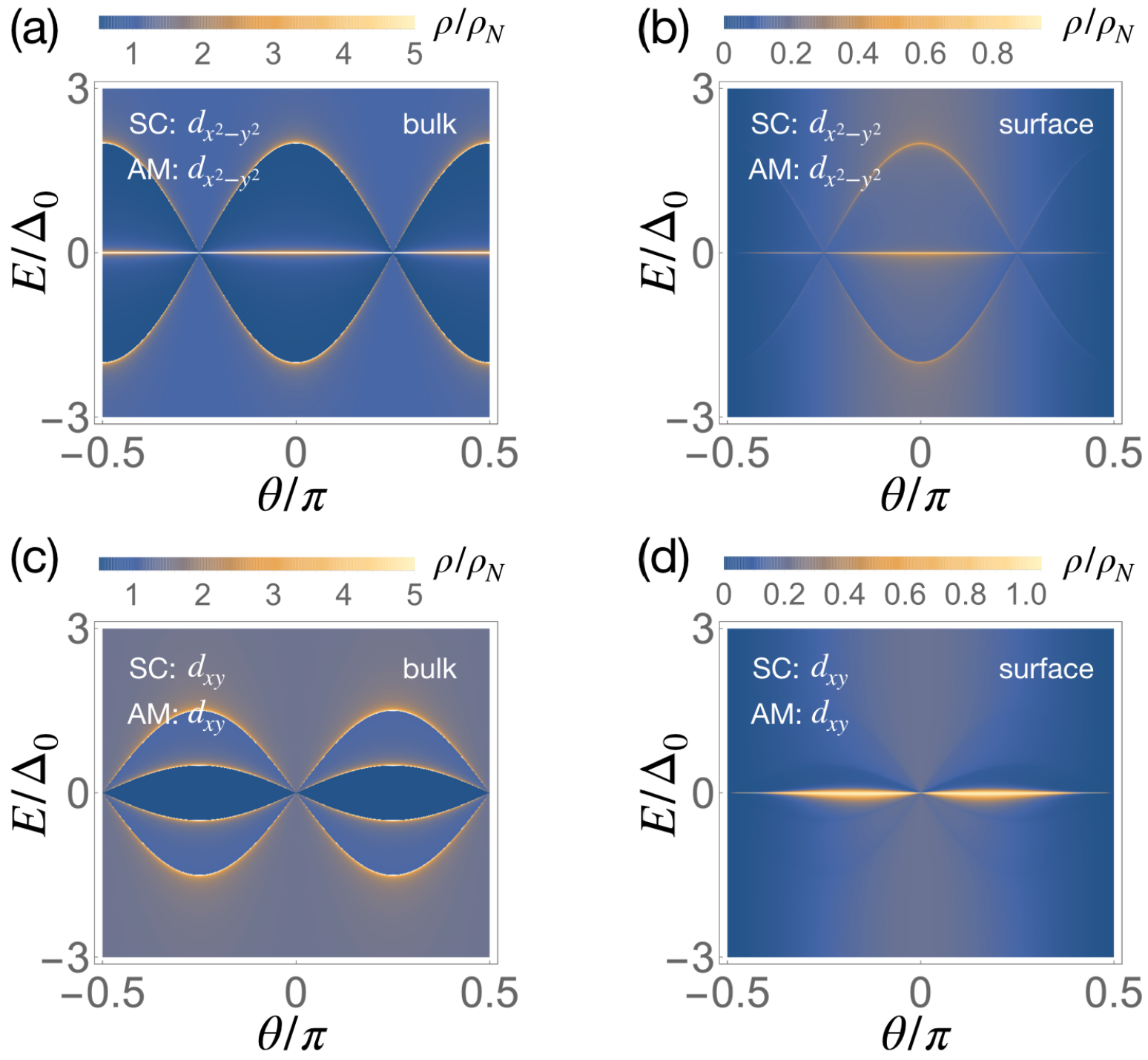}
\end{center}
\caption{ Angle-resolved density of states $\rho$ for junctions with $d$-wave SCs. Both SC and AM are $d_{x^{2}-y^{2}}$-wave in (a)(b) and $d_{xy}$-wave in (c)(d). (a)(c) are in the bulk while (b)(d) are at the surface $x=0$ with barrier strength $Z=2$. The AM strength is $h_0=\Delta_0$ for (a)(b) and $h_0=0.5\Delta_0$ for (c)(d).  $\rho_N$ is the angle-resolved density of states in the normal state.}
\label{dosdwave}
\end{figure}

\begin{figure}[t]
\begin{center}
\includegraphics[width=0.6\linewidth]{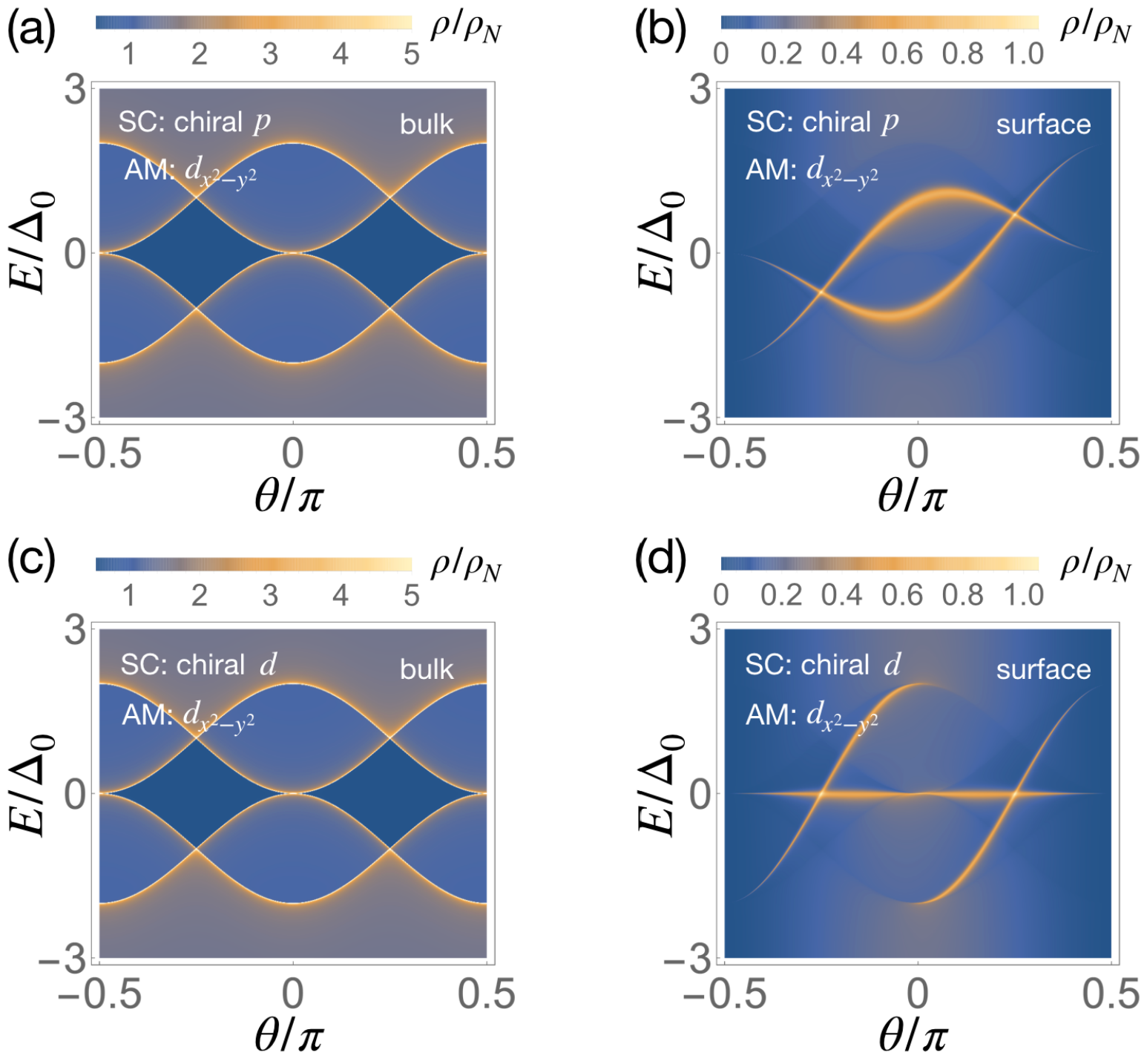}
\end{center}
\caption{Angle-resolved density of states for junctions with chiral SCs. The SC is chiral $p$-wave in (a)(b) and chiral $d$-wave in (c)(d). AM are $d_{x^2-y^2}$ for all panels. (a)(c) are in the bulk while (b)(d) are at the surface $x=0$ with barrier strength $Z=2$. The AM strength is $h_0=\Delta_0$ for all panels.}
\label{doschiral}
\end{figure}

\noindent \textbf{\textbf{S7.} Surface and bulk density of states in normal-superconductor junctions} \newline

In this section, we address the surface density of states as well as bulk
density of states based on McMillan's Green's function \cite{BoMcMillan}. We can construct the
Green's function of the normal-superconductor junction in the main text by
scattering waves and obtain the local density of states $\rho$. We will show the calculation details elsewhere but present
the main results here. From Fig.\ref{dosdwave}(a,b), we can clearly see the
enhanced zero-energy density of states both in the bulk and at the surface
for a $d_{x^{2}-y^{2}}$-wave SC with a $d_{x^{2}-y^{2}}$-wave AM. We thus
can conclude that these flat bands are owing to the bulk effect, not the
boudary effect due to their universal existence. On the opposite, Fig.\ref%
{dosdwave}(c,d) show a standard example of surface states from our previous
results for a $d_{xy}$-wave SC with a $d_{xy}$-wave AM. It is seen that the
bulk density of states and surface density of states behave differently,
consistent with the result of SABSs since the surface bound
states only reside close to the boundary but disappear deep into the bulk.
For chiral wave superconductors, we show both bulk and surface density of
states in Fig.\ref{doschiral}. As seen in the Figures, the bulk and surface
states are distinct while the chiral states show up at the surface which are
consistent with the main text.